\documentclass[twocolumn]{revtex4-1}
\pdfoutput=1
\usepackage[utf8]{inputenc}
\usepackage[english]{babel}
\usepackage[T1]{fontenc}
\usepackage{tikz}
\usepackage{lipsum}
\usepackage{amssymb}
\usepackage{amsmath}
\usepackage{amsbsy}
\usepackage{mathtools}
\usepackage{mathrsfs}
\usepackage{physics}
\usepackage{mathtools}
\usepackage{mathrsfs}
\usepackage{physics}

\usepackage[colorlinks=true, allcolors=blue]{hyperref}

\newcommand{\uama}{\'Area de F\'isica Te\'orica y
Materia Condensada, Universidad Aut\'onoma Metropolitana
Azcapotzalco, Av. San Pablo 180, Col. Reynosa-Tamaulipas,
02200 Cuidad de M\'exico, M\'exico}
\newcommand{\ifunam}{Instituto de F\'isica,
Universidad Nacional Aut\'onoma de M\'exico,
Apartado Postal 20-364, M\'exico, Distrito Federal 01000, M\'exico}

\newcommand{\ve}[1]{\boldsymbol{#1}}

\begin{document}

\title{Vectorization of the density matrix and quantum
simulation of the von Neumann equation
of time-dependent Hamiltonians.
}

\author{Alejandro Kunold }
\affiliation{ \uama}
\affiliation{ \ifunam}
\email{akb@azc.uam.mx}

\begin{abstract}
Based oh the properties of Lie algebras,
in this work we develop a general framework
to linearize the von-Neumann equation
rendering it in a suitable form
for quantum simulations.
We show that one of these linearizations
of the von-Neumann equation
corresponds to the standard case
in which the state vector becomes
the column stacked elements of the density
matrix and the Hamiltonian superoperator takes
the form $I\otimes H-H^\top \otimes I$
where $I$ is the identity matrix and $H$
is the standard Hamiltonian.
It is proven that this particular form
belongs to a wider class
of ways of linearizing
the von Neumann equation
that can be categorized by the algebra from
which they originated.
Particular attention is payed to
Hermitian algebras that
yield real density matrix coefficients
substantially simplifying the 
quantum tomography of the state vector.
Based on this ideas, a quantum algorithm to
simulate the dynamics of the density matrix
is proposed.
It is shown that this method, along
with the unique properties of the algebra formed by Pauli
strings allows to avoid the use of Trotterization
hence considerably reducing the circuit depth.
Even though we have used the special case
of the
algebra formed by the Pauli strings,
the algorithm
can be readily adapted to other algebras.
The algorithm is demonstrated for two
toy Hamiltonians using the IBM
noisy quantum circuit
simulator.

\end{abstract}

\maketitle

\section{Introduction}
One of the most important applications
of quantum computers is the simulation
of large quantum systems
that are intractable
using conventional classical computers
\cite{feynman1982simulating, lloyd1996universal,
nielsen2002quantum}.
A vast number of quantum techniques 
to simulate quantum dynamics has been developed
with this purpose in mind
\cite{RevModPhys.86.153, miessen2023quantum}.
Specially many body systems
give rise to extremely large Hamiltonians
\cite{PhysRevLett.79.2586}
that are classically impossible to tackle even
for a small number of particles.

A very common quantum simulation problem
is the digital quantum simulation (DQS)
of a closed quantum system
\cite{lloyd1996universal,
PhysRevLett.79.2586, PhysRevE.56.3661,
Zalka1998, PhysRevA.64.022319,MARZUOLI200279,
PhysRevA.65.042323,PhysRevA.79.032316,
berry2007efficient,
Raeisi_2012,RevModPhys.86.153}
which consists in solving
the time-dependent Schr\"odinger
equation.
In general terms, a DQS algorithm
consists of two steps. First the
time evolution operator of a
given Hamiltonian is expressed as
a series of unitary quantum gates
through the Jordan-Wigner isomorphism
\cite{PhysRevA.65.042323}, for example.
Second, the time evolution operator is used to
propagate an initial
state \cite{lloyd1996universal, miessen2023quantum},
typically by shifting locally the wave function
forward in time over discrete and
sufficiently small time slices
\cite{lloyd1996universal,RevModPhys.86.153,miessen2023quantum}.
The Hamiltonian is written as the sum over many
simpler interactions
$H=\sum_{i=1}^mH_i$ and the evolution operator
can be approximated by
$\mathcal{U}(t)=\mathrm{e}^{-iH_1\Delta t/\hbar}
\mathrm{e}^{-iH_2\Delta t/\hbar}
\dots \mathrm{e}^{-iH_m \Delta t/\hbar}$
as long as the interval $\Delta t$ is small enough.
Through the Trotter-Suzuki formula
\cite{trotter1958approximation,SUZUKI1992387,10.1063/1.4952761},
that takes into account the non-commutativity of
the $H_i$'s,
the accuracy of $U(t)$
can then be improved but at the cost of
increasing the circuit depth.

Nearly all realistic quantum systems
are open systems whose evolution is non-unitary
due to the decoherence induced by the
unavoidable interaction with the environment.
This presents a major difficulty
for digital quantum computers
that can only operate with a set of unitary
gates.
This obstacle has spawned
a wide range of solutions
for the quantum simulation of open
quantum systems
\cite{PhysRevA.62.032309,
di2015quantum, PhysRevA.83.062317,
wei2016duality,
motta2020determining,
hu2020quantum,
PRXQuantum.3.010320}.
The following are some examples.
The open-system nature
of a two-spin NMR ensemble
has been employed to simulate
the decoherence in a quantum simulation
\cite{PhysRevA.62.032309}.
Also, a set of ancilla qubits that
are designed to have the same effect as
the simulated environment
have been used to represent the
decoherence of an open quantum
system \cite{PhysRevA.83.062317}.
Quantum algorithms
have been developed specifically
for duality quantum computers
that can perform non-unitary operations
\cite{wei2016duality}.
More recently,
eigenstate and thermal state quantum simulations
have been demonstrated employing
the quantum analogues
of imaginary time evolution
and Lanczos algorithms
\cite{motta2020determining}.
Of direct relevance to this work
is a new kind of open-quantum system simulation
algorithm that solves the Lindblad
master equation
and that has successfully been tested
in real digital quantum hardware \cite{PRXQuantum.3.010320}.
The algorithm mainly
relies on
the Kraus matrix representations
\cite{KRAUS1971311, havel2003, PhysRevA.74.062113}
and an adaptation of
the imaginary time evolution algorithm \cite{motta2020determining}.
In general, the Lindblad equation
is first vectorized in the form of a Scrh\"odinger
equation through the Kraus operator sum
\cite{KRAUS1971311, havel2003, 
PhysRevA.74.062113,Ramusat2021quantumalgorithm}
and then, the anti-Hermitian component of
the Hamiltonian is treated through
quantum imaginary evolution.
The elements of the vectorized density matrix
are finally obtained by quantum tomography.
The unitary evolution of the Hermitian component
of the Hamiltonian is calculated
by the standard techniques
of DQS listed above.

In this work we are primarily interested in the
linearization of the density matrix
that is governed by the von Neumann equation
\begin{equation}
i\hbar\frac{d}{dt}\rho(t) =  \left[H(t), \rho(t)\right],
\label{eq:vonneumanneq}
\end{equation}
where $H$ and $\rho(t)$ are the
$N\times N$ matrices for the Hamiltonian and the
density matrix.
The linearization is typically done
by expressing the density matrix
using the Kraus operator sum and
column stacking
the density matrix elements.
As we show below,
this procedure yields the following linearized
von Neumann equation \cite{havel2003,PRXQuantum.3.010320}
\begin{equation}
i\hbar\frac{d}{dt}\left\vert \boldsymbol{\rho} \right\rangle
=\left[I\otimes H(t)-H^\top(t)\otimes I\right]
\left\vert \boldsymbol{\rho} \right\rangle,
\label{eq:stackedvonneumann}
\end{equation}
where $I$ is the $N\times N$ identity matrix and
$\left\vert \boldsymbol{\rho} \right\rangle$
denotes the vector resulting
from stacking the columns
of $\rho(t)$ from left to right.

We develop a method that generalizes the
vectorization
process of the von Neumann equation
expanding the density matrix in terms
of the elements of a Lie algebra.
Moreover, we show that the column-stacked
density matrix in Eq. \eqref{eq:stackedvonneumann}
is in fact one of the
many possible ways of
representing $\rho(t)$
as a vector using
the elements of a very
specific Lie algebra.
Instead of using this algebra,
whose elements are non Hermitian,
we propose expanding the
von Neumann equation in terms
of Hermitian matrices.
The advantage is that the thus vectorized density matrix
has purely real coefficients that
greatly simplify the quantum tomography of
$\left\vert \boldsymbol{\rho} \right\rangle$.
We present the special case of
the algebra formed by Pauli strings and
with it, implement an algorithm
that solves the von Neumann equation
to illustrate the method.
The method, together with the special properties of
Pauli strings, allows to
avoid the use of Trotterization
therefore reducing the circuit depth.
The algorithm was tested for the
von Neumann equation of the
magnetic resonance Hamiltonian
using Qiskit in the noisy quantum circuit
simulator {\it qasm\_simulator} \cite{ibm} .
The code can be
downloaded from \cite{GitHub}.

\section{Vectorization of the density matrix. }
Any $N\times N$ square matrix $W$ can be 
expressed as the linear combination
of the elements of the finite Lie algebra
$\mathfrak{g}_n = \left\{g_1, g_2, \dots, g_n\right\}$,
as
\begin{equation}
W = \sum_{i=1}^n w_i g_i = \ve{g}^\top\ve{w}
\label{eq:wexpan}
\end{equation}
where $\ve{g}^\top=\left(g_1, g_2,\dots,g_n\right)$,
$\ve{w}^\top=\left(w_1, w_2,\dots,w_n\right)$,
$g_i \in \mathbb{C}^{N\times N}$ and $w_i \in \mathbb{C}$.
If $W$ is known, the coefficients $w_i$
can be calculated as 
\begin{equation}
w_i =\frac{1}{\mathrm{Tr}\left[g^\dagger_ig_i\right]}
\mathrm{Tr}\left[g^\dagger_i W\right],
\label{eq:wproj}
\end{equation}
provided that the elements of the algebra
are orthogonal under the Frobenius inner
product, namely
\begin{equation}
\mathrm{Tr}\left[g_i^\dagger g_j\right]
=\mathrm{Tr}\left[g^\dagger_ig_i\right]\delta_{i,j}.
\label{eq:frobenius}
\end{equation}
Not all algebras have orthogonal elements,
however, one can always find a linear combination
of them that
does meet \eqref{eq:frobenius}.
Such combinations might be generated, for example,
through the Gram-Schmidt method.

One of the fundamental properties of Lie
algebras is that their elements
form a closed commutator algebra
described by
\begin{equation}
    \left[g_i, g_j\right] = i\hbar\sum_k c_{i,j,k}g_k.
    \label{eq:commug}
\end{equation}
where the Lie bracket of $\mathfrak{g}_n$
has been chosen to be the commutator $\left[g_i, g_j\right]=g_ig_j-g_jg_i$.
From this definition it is straight forward
to show that the structure constants
$c_{i,j,k}$ have the property
\begin{equation}
c_{i,j,k}=-c_{j,i,k}.
\label{eq:struconsprop1}
\end{equation}
The explicit form of the structure constants
\begin{equation}
c_{i,j,k} =\frac{ \mathrm{Tr}\left[g^\dagger_k\left[g_i, g_j\right]\right]}
{{i\hbar}\Tr[g_k^\dagger g_k]}
=\frac{ \mathrm{Tr}\left[\left[g_i, g_j\right]g^\dagger_k\right]}
{{i\hbar}\Tr[g_k^\dagger g_k]}
\label{eq:scprop1}
\end{equation}
is obtained by
combining Eqs. \eqref{eq:frobenius}
and \eqref{eq:commug}.
In the case of an algebra
$\mathfrak{h}_n=\left\{h_1, h_2, \dots, h_n\right\}$
formed only
by Hermitian matrices the structure constants
acquire a new and useful property,
the invariance with respect
to cyclic permutation of the indices
\begin{multline}
c_{i,j,k}=\frac{\mathrm{Tr}\left[\left[h_i,h_j\right]h_k\right]}
{i\hbar \mathrm{Tr}\left[h_kh_k\right] }
=\frac{\mathrm{Tr}\left[h_ih_jh_k-h_jh_ih_k\right]}
{i\hbar \mathrm{Tr}\left[h_kh_k\right] }\\
=\frac{\mathrm{Tr}\left[h_kh_ih_j-h_ih_kh_j\right]}
{i\hbar \mathrm{Tr}\left[h_kh_k\right] }
=\frac{\mathrm{Tr}\left[\left[h_k,h_i\right]h_j\right]}
{i\hbar \mathrm{Tr}\left[h_ih_i\right] }
=c_{k,i,j},
\label{eq:scprop2}
\end{multline}
as long as $\mathrm{Tr}\left[h_ih_i\right]=\mathrm{Tr}\left[h_kh_k\right]$.
Note that the properties
\eqref{eq:scprop1} and \eqref{eq:scprop2}
imply that $c_{i,j,k}$
is fully antisymmetric.

With these definitions at hand we can proceed to
vectorize the von Neumann equation.
In terms of the most general algebra $\mathfrak{g}_n$
the density matrix can be expanded as
\begin{equation}
\rho(t) = \sum_i g_i \rho_i(t) = \ve{g}^\top \ve{\rho}(t) ,
\end{equation}
where $\ve{\rho}(t)=(\rho_1(t), \rho_2(t), \dots, \rho_n(t))$
and from Eq. \eqref{eq:wproj} the coefficients are
\begin{equation}
\rho_i(t) = \frac{1}{\Tr\left[g_i^\dagger g_i\right]}
\left[g_i^\dagger \rho(t)\right].
\end{equation}
Multiplying
the left and right-hand side terms of \eqref{eq:vonneumanneq}
by $g_k^\dagger$ and tracing
it follows that
\begin{multline}
i\hbar \frac{d}{dt}\Tr\left[g_k^\dagger \rho (t)\right]
=\sum_j\Tr\left[g_k^\dagger \left[H(t),g_j\right]\right]
\Tr[g^\dagger_j\rho(t)]\\
=\sum_{i,j}\Tr[g^\dagger_ig_i]a_i(t)\Tr\left[g_k^\dagger \left[g_i,g_j\right]\right]
\Tr[g^\dagger_j\rho(t)]\\
=i\hbar\sum_{i,j}\Tr[g^\dagger_kg_k]\Tr[g^\dagger_ig_i]
c_{i,j,k}a_i(t)\Tr[g^\dagger_j\rho(t)].
\label{eq:expanvng}
\end{multline}
Because of \eqref{eq:wexpan} and \eqref{eq:wproj},
the Hamiltonian can be expressed as the
decomposition
\begin{equation}
H(t)=\sum_i a_i(t)g_i=\ve{a}^\top(t) \ve{g}
\end{equation}
where
\begin{equation}
a_i(t) =\frac{1}{\mathrm{Tr}\left[g_i^\dagger g_i\right]}
\mathrm{Tr}\left[g_i^\dagger H(t)\right],
\label{eq:hamprojectiong}
\end{equation}
are the Hamiltonian coefficients
and $\ve{a}^\top(t) = (a_1(t), a_2(t), \dots, a_n(t))$.
Equation \eqref{eq:expanvng}
can be put in the more succinct form
\begin{equation}
i\hbar\frac{d}{dt}\ve{\rho}(t) =  \mathcal{H}(t)\ve{\rho}(t)
\label{eq:linvonneu}
\end{equation}
where
\begin{equation}
\mathcal{H}(t)=\sum_i\mathcal{H}_i,
\label{eq:superhamg}
\end{equation}
is the Hamiltonian superoperator,
\begin{equation}
\mathcal{H}_i=i\hbar\Tr[g^\dagger_kg_k]\Tr[g^\dagger_ig_i]
C_{i}a_i(t),
\end{equation}
and $C_i$ is the matrix whose elements are 
the structure constants $(C_i)_{j,k}=c_{i,j,k}$.
Equation \eqref{eq:linvonneu} is a
general form of the vectorized
von Neumann equation.

At this point it is natural to ask what is
the connection between this version of the
von Neumann equation and \eqref{eq:stackedvonneumann}.
To address this question we introduce two
auxiliary algebras.
First consider the algebra
$\mathfrak{q}_n=\left\{q_1, q_2, \dots, q_n\right\}$
formed by the $N\times N$ sparse matrices $q_i$ that
only have a $1$ that progressively shifts
from top to bottom and from left to right as $i$ goes
from $1$ to $n=N^2$.
More explicitly
$(q_i)_{j,k}=\delta_{j,i_1}\delta_{k,i_2}$
where
\begin{eqnarray}
i_1 &=& (i-1)\bmod N +1, \label{eq:i1}\\
i_2 &=& [(i-1)/N]+1.\label{eq:i2}
\end{eqnarray}
Therefore, the vector of dimension $n=N^2$
formed by the coefficients of a given
$N\times N$  matrix in terms of $\mathfrak{q}_n$
corresponds to the stacked columns thereof.
Indeed, the vector formed by the coefficients of $W$
in terms of
$\mathfrak{q}_n$ are the matrix elements of $W$
\begin{equation} 
w_i=\frac{\Tr[q_i^\dagger W]}{\Tr[q_i^\dagger q_i]}
= \Tr[q_i^\dagger W]=(W)_{i_1,i_2}
\end{equation}
where $i_1$ and $i_2$ are given by \eqref{eq:i1} and
\eqref{eq:i2}, respectively.

The second auxiliary algebra is
$\mathfrak{q}_n^{\otimes 2} $
$=\mathfrak{q}_n\otimes \mathfrak{q}_n $
$=\{q_1\otimes q_1,$ $ q_1\otimes q_2,$
$\dots, q_{n-1}\otimes q_{n},$ $ q_n\otimes q_n \}$
$=\{Q_{1,1},$ $ Q_{1,2},$ $ \dots, Q_{n,n-1},$ $ Q_{n,n}\}$
where $Q_{i,j}$ is an $N^2\times N^2$ sparse matrix
that has a $1$ in the $i,j$ position. It is important
to note that even though in general $Q_{i,j}\ne q_i\otimes q_j$,
it is possible to establish a one to one relation
\begin{equation}
Q_{k,l} = q_b\otimes q_a
\label{eq:Qqqrel1}
\end{equation}
for
\begin{eqnarray}
a &=& N \left[\left(k-1\right)\bmod N\right]
+ \left(l-1\right)\bmod N + 1,
\label{eq:Qqqrel2}\\
b &=& N \left[\left(k-1\right)/ N\right]
+ \left[\left(l-1\right)/N\right] + 1.
\label{eq:Qqqrel3}
\end{eqnarray}

Following a similar procedure as in Eq. \eqref{eq:expanvng}
the expansion of the von Neumann equation in the $\mathfrak{q}_n$
altakes the form
\begin{multline}
i\hbar \frac{d}{dt}\Tr\left[q_k^\dagger \rho (t)\right]
=\sum_j\Tr\left[q_k^\dagger \left[H(t),q_j\right]\right]
\Tr[q^\dagger_j\rho(t)]\\
=\sum_j\Tr\left[q_k^\dagger H(t)q_j I-q_k^\dagger I q_jH(t)\right]
\Tr[q^\dagger_j\rho(t)].
\label{eq:vectorvn1}
\end{multline}
In Appendix \ref{ap:Qiden} it is shown that
if $A$ and $B$ are two $N\times N$ matrices then
\begin{equation}
\Tr\left[q_i^\dagger A q_j B\right]
=\Tr\left[Q_{i,j}^\dagger B^\top \otimes A\right]
=(B^\top \otimes A)_{i,j}.
\end{equation}
With this, Eq. \eqref{eq:vectorvn1} becomes
\begin{multline}
i\hbar \frac{d}{dt}\Tr\left[q_k^\dagger \rho (t)\right]\\
=\sum_j \bigg(
I\otimes H(t)-H^\top(t) \otimes I\bigg)_{k,j}
\Tr\left[q_j^\dagger \rho (t)\right],
\label{eq:vectorvn2}
\end{multline}
that,
in full consistency with Eq. \eqref{eq:stackedvonneumann},
can also be written as the vectorial expression
\begin{equation}
i\hbar \frac{d}{dt}\ve{\rho}(t)
=\sum_j \bigg(
I\otimes H(t)-H^\top(t) \otimes I\bigg)
\ve{\rho}(t).
\label{eq:vectorvn2}
\end{equation}
Hence, the von Neumann equation that stems
from the Kraus sum is a particular case of
\eqref{eq:linvonneu} when it is expanded
in terms of the algebra $\mathfrak{q}_n$.
In this case $\ve{\rho}(t)$ corresponds to the
stacked columns of the density matrix.
The entries of the vector $\ve{\rho}(t)$ are, however,
complex numbers since $\mathfrak{q}_n$ is comprised
of non-Hermitian matrices. This substantially
complicates the quantum tomography of $\ve{\rho}(t)$
whose elements may have non-vanishing phases.

To circumvent this difficulty one can
use any Hermitian algebra $\mathfrak{h}_n$
that yields real entries for $\ve{\rho}(t)$.
In this base, the Hamiltonian and the density matrix
\begin{eqnarray}
H(t) = \sum_i a_i(t) h_i = \boldsymbol{a}^\top(t) \boldsymbol{h},
\label{eq:ham}\\
\rho(t)=\sum_{i=1}^n h_i\rho_i(t) =\ve{h}^\top\ve{\rho}(t) 
\label{eq:evolrho1}
\end{eqnarray}
are expressed as the linear combination
of the elements of the finite Lie algebra
$\mathfrak{h}_n = \left\{h_1, h_2, \dots, h_n\right\}$
with $\boldsymbol{h}^\top=(h_1, h_2, \dots, h_n)$ and $n=2^N$.
The Hamiltonian coefficients
\begin{equation}
a_i(t) =\frac{1}{\mathrm{Tr}\left[h_ih_i\right]}
\mathrm{Tr}\left[H h_i\right].
\label{eq:hamprojectionh}
\end{equation}
are real and, in general, time-dependent.
Similarly, the density matrix coefficients
are
\begin{equation}
\rho_i(t) =\frac{1}{\mathrm{Tr}\left[h_ih_i\right]}
\mathrm{Tr}\left[\rho(t) h_i\right].
\label{eq:rhoprojectionh}
\end{equation}
As before, $\ve{a}(t)=(a_1(t), a_2(t), \dots, a_n(t))$
and $\ve{\rho}(t)=(\rho_1(t), \rho_2(t), \dots, \rho_n(t))$.
From \eqref{eq:superhamg} the Hamiltonian superoperator
is given by
\begin{equation}
\mathcal{H}(t)=\sum_i\mathcal{H}_i(t)
\end{equation}
where
\begin{equation}
\mathcal{H}_{i}(t)=i\hbar\Tr[h_k h_k]\Tr[h_i h_i]
C_{i}a_i(t).
\end{equation}
Note that $\mathcal{H}$ is Hermitian
because the coefficients $a_i(t)$
are real and
the structure constants
are fully antisymmetric, i.e.,
\begin{equation}
c_{i,j,k}=-c_{i,k,j},
\label{eq:strucons1}
\end{equation}
and therefore
$C_k=-C_k^\top$ is skew-symmetric.

\begin{figure*}
  \includegraphics[width=0.950\textwidth]{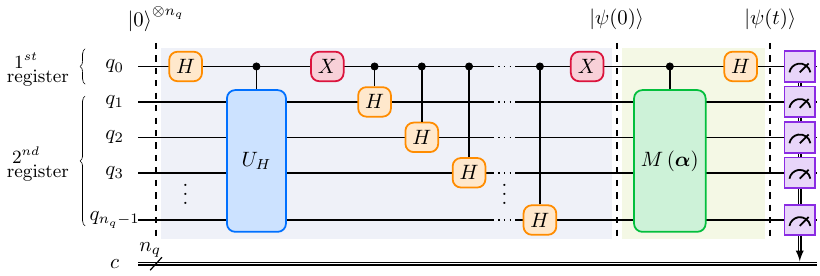}
  \caption{
    Main quantum algorithm to determine the dynamics of the density
    matrix coefficients $\rho_1(t)$,$\dots$, $\rho_n(t)$. 
    (b) Controlled $M\left(\ve{\alpha}\right)$ gate expressed
    as a sequence of differential time steps.
    (c) One time step controlled differential $M\left(d\ve{\alpha}\right)$
    gate expressed as the series of Hamiltonian gates generated by
    the structure constants of the Pauli strings.
    }
    \label{fig01}
\end{figure*}

\section{Time evolution}
At first glance it would seem reasonable
to find the evolution operator
by directly applying
the standard techniques of DQS
to \eqref{eq:linvonneu}.
However, to do so efficiently,
it would be convenient to
leverage the structure and
the symmetries of
the linear operator $\mathcal{H}$
in \eqref{eq:superhamg}.
These are not obvious from the
expression of $\mathcal{H}$.
But the contraction $C_{i}a_i(t)$ ($i= 1,2,\dots, n$)
signals the fact that only the $n$
structure constants are needed to generate the time
evolution \cite{10.1063/1.4952761}
and not $n^2 =n\times n$ as the dimension
of $\mathcal{H}$ would suggest.
Instead of following this line of reasoning
we resort to the Lie algebraic method
to prove that indeed only $n$ elements
are needed to generate the time evolution.

The Lie algebraic method
enables the exact determination of evolution operators
for time-dependent Hamiltonians of finite dimension
or having a dynamical algebra.
Though the general method is thoroughly discussed in
\cite{annphys531} a quick review
is provided here for completeness.

\begin{figure*}
  \includegraphics[width=0.950\textwidth]{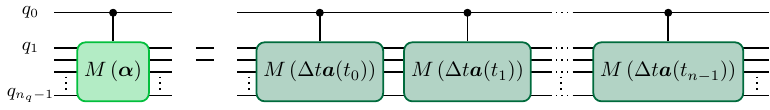}
  \caption{
    Controlled $M\left(\ve{\alpha}\right)$ gate expressed
    as a sequence of differential time steps.
    }
    \label{fig02}
\end{figure*}

One possible way of expressing the evolution operator
of the Hamiltonian \eqref{eq:ham} is
$\mathcal{U}(t)=U^{\dagger}(t)
= U_1^\dagger(t)U_2^\dagger(t) \dots U_n^\dagger(t)$
where
\begin{equation}
U_k(t)=\exp\left[\frac{i}{\hbar} \alpha_k(t) h_k\right],
\label{eq:ukdef}
\end{equation}
is the unitary transformation generated by
the $k$-th element of $\mathfrak{h}_n$.
Each $U_k$ transforms $\boldsymbol{h}$ according to
\begin{equation}
U_k(t)\boldsymbol{h}U_k^\dagger(t) =
M_k\left(\alpha_k(t)\right)\boldsymbol{h},
\end{equation}
where the explicit form of the
matrices $M_k\left(\alpha_k(t)\right)$
is given by
\begin{equation}
M_k(\alpha_k(t)) = \exp\left[-\alpha_k(t) C_k\right].
\label{eq:mkgate}
\end{equation}
The matrices $M_k(\alpha_k(t))$ are unitary
since, as mentioned earlier, $C_k=-C_k^\top$ is skew-symmetric 
in accordance with \eqref{eq:strucons1}.
Therefore, by performing a time evolution
over $\ve{h}$ we obtain
\begin{equation}
    U(t)\ve{h}U^{\dagger}(t)=M(\ve{\alpha}(t))\ve{h},
    \label{eq:evolh}
\end{equation}
where
\begin{equation}
M(\ve{\alpha}(t)) = M_1(\alpha_1(t))M_2(\alpha_2(t))
  \dots M_n(\alpha_n(t)).
\label{eq:mexact}
\end{equation}

The explicit time-dependence of the $\alpha_k(t)$ parameters
is yet unknown but can be derived from the Schr\"odinger equation
\begin{equation}
\big(H(t) - p_t\big)\left\vert \psi(t)\right\rangle = 0,
\end{equation}
where $p_t=i\hbar\partial/\partial t$ is the energy operator.
The Hamiltonian and the energy operator transform according to
\begin{eqnarray}
U(t) H U^\dagger(t) 
&=& \boldsymbol{a}^\top M_1 M_2\dots M_n \boldsymbol{h}
= \boldsymbol{a}^\top M \boldsymbol{h}, \label{eq:transham}\\
U(t) p_t U(t)^\dagger 
 &=& p_t + \dot{\boldsymbol{\alpha}}^\top V^\top \boldsymbol{h}
 \label{eq:transpt}
\end{eqnarray}
where
\begin{eqnarray}
V^\top &=& I_1 M_2(\alpha_2) \dots M_n(\alpha_n)
     \nonumber\\
      &+& I_2 M_3(\alpha_3) \dots M_n(\alpha_n)
       + \dots + I_n ,
       \label{eq:vexact}\\
(I_k)_{ij} &=& \delta_{i,k}\delta_{j,k}.
\label{eq:defI}
\end{eqnarray}
It then follows from Eqs. \eqref{eq:transham} and \eqref{eq:transpt}
that the Schr\"odinger equation under
$U(t)$ becomes
\begin{multline}
U(t)\left(H-p_t\right)U^\dagger(t) U(t)\left\vert \psi(t) \right\rangle \\
=\left[\boldsymbol{a}^\top M \boldsymbol{h}
- \dot{\boldsymbol{\alpha}}^\top V^\top\boldsymbol{h}-p_t\right]
U(t)\left\vert \psi(t)\right\rangle =0.
\end{multline}
Making
\begin{equation}
\ve{a}^\top(t) M\left(\ve{\alpha}(t)\right) \ve{h}
- \dot{\ve{\alpha}}(t)^\top V^\top\left(\ve{\alpha}(t)\right)
   \ve{h} = 0,
\end{equation}
or, equivalently
\begin{equation}
\dot{\ve{\alpha}}(t)
=\ve{F}\left(t,\boldsymbol{\alpha}(t)\right)
= V^{-1}\left(\ve{\alpha}(t)\right)
M^\top\left(\ve{\alpha}(t)\right) \ve{a}(t),
\label{eq:alphadotexact}
\end{equation}
we arrive at the result that
$p_t U(t)\left\vert \psi(t)\right\rangle =0$.
Since $p_t=i\hbar\partial/\partial t$, the previous
equation will only hold if $\mathcal{U}(t)=U^\dagger(t)$ and
$U(t)\left\vert \psi(t)\right\rangle=\left\vert \psi_0\right\rangle
=\left\vert \psi(0)\right\rangle$
is the initial state.
Equation \eqref{eq:alphadotexact} consists
of a system of $n$ ordinary differential equations
for $\alpha_1(t), \alpha_2(t)$, $\dots \alpha_n(t)$.
The explicit time dependence of the $\ve{\alpha}(t)$
parameters comes from
the solution of \eqref{eq:alphadotexact} along with
the initial condition 
\begin{equation}
\ve{\alpha}(0)=0,
\label{eq:incondalpha}
\end{equation}
necessary to ensure that $U(0)=\mathcal{U}(0) = 1$.

As we mentioned above, since both $\rho(t)$ and $\ve{h}$ 
are Hermitian matrices, the elements of $\ve{\rho}$
are real. This crucial feature
significantly simplify the quantum tomography
of the density matrix coefficients
allowing to
easily determine its elements
through phase kickback \cite{cleve1998quantum}
with only one auxiliary qbit.
Using Eqs. \eqref{eq:evolh} and \eqref{eq:evolrho1}
the evolution of the density matrix operator in the Schr\"odinger picture
is expressed according to
\begin{multline}
\rho(t)=\mathcal{U}(t)\rho(0)\mathcal{U}^\dagger(t)
=U^\dagger(t)\rho(0)U(t)\\
=\sum_iU^\dagger(t)h_iU(t)\rho_i(0)
=U^\dagger(t)\ve{h}^\top U(t)\ve{\rho}(0)\\
=\ve{h}^\top(0)M(\ve{\alpha}(t))\ve{\rho}(0).
\end{multline}
From the comparison of this result
with Eq. \eqref{eq:evolrho1}
it follows 
that the time-dependent density matrix may be
cast in the shape of a vector as
\begin{equation}
\ve{\rho}(t)=M(\ve{\alpha}(t))\ve{\rho}(0).
\label{eq:coreeq1}
\end{equation}
Except for a normalization constant, notice that
$M^\top(\ve{\alpha}(t))$ is associated to
the superoperator of $U^\dagger(t)$ acting on the
vectorized density matrix $\ve{\rho}(t)$
in the Fock-Liouville space generated by $\mathfrak{h}_n$.
In other words,
the evolution of the density matrix thus
maps onto the evolution of a
quantum state vector
\begin{equation}
\left\vert\ve{\rho}(t)\right\rangle =
M(\ve{\alpha}(t))\left\vert\ve{\rho}(0)\right\rangle,
\end{equation}
where $M(\ve{\alpha}(t))$ is a super-evolution operator.
The density matrix can be expanded as
\begin{equation}
\left\vert \boldsymbol{\rho}(t)\right\rangle
=\sum_i  \rho_i(t)\left\vert h_i \right\rangle,
\end{equation}
where $\rho_i(t)$ are the normalized density matrix
coefficients ($\sum_i\rho_i^2(t)=1$) and there
is a one-to-one relation between the
Fock-Liouville states $\left\vert h_i\right\rangle$
and the matrices $h_i$.

\begin{figure*}
  \includegraphics[width=0.950\textwidth]{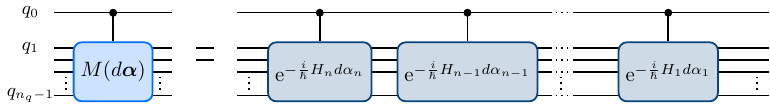}
  \caption{
    One time step controlled differential $M\left(d\ve{\alpha}\right)$
    gate expressed as the series of Hamiltonian gates generated by
    the structure constants of the Pauli strings.
    }
    \label{fig03}
\end{figure*}

Equation \eqref{eq:coreeq1} proves that the time
evolution of the vectorized density matrix
is generated by only $n$ elements
that correspond to the structure constants $C_k$
as was hinted at the beginning of this section.

\section{Algorithm}
We now provide a quantum simulation algorithm for the density
matrix $\rho(t)$ obeying the von-Neumann equation
\eqref{eq:vonneumanneq} expanded in a
general algebra $\mathfrak{h}_n$.
It illustrates the time evolution
and the quantum tomography of the density matrix.

Equation \eqref{eq:coreeq1} is the underpinning
element of the algorithm. It allows
for the computation of $\rho(t)$ through a series
of time steps $t_m$ of step size $\Delta t=t_m-t_{m-1}$ as
\begin{equation}
\left\vert \ve{\rho}(t_m)\right\rangle = 
  M(d\ve{\alpha}_m)
  M(d\ve{\alpha}_{m-1})
  \dots
  M(d\ve{\alpha}_{1})
  \left\vert \ve{\rho}(0)\right\rangle,
  \label{eq:rhotimesteps}
\end{equation}
where, from \eqref{eq:alphadotexact}
\begin{multline}
d\ve{\alpha}_m=\ve{\alpha}(t_m)-\ve{\alpha}(t_{m-1})
=\frac{\dot{\ve{\alpha}}(t_{m-1})}{1!}\Delta t\\
+\frac{\ddot{\ve{\alpha}}(t_{m-1})}{2!}\Delta t^2
+\dots
=\ve{F}\left(t_{m-1},\boldsymbol{\alpha}(t_{m-1})\right)
\frac{\Delta t}{1!}\\
+\dot{\ve{F}}\left(t_{m-1},\boldsymbol{\alpha}(t_{m-1})\right)
\frac{\Delta t^2}{2!}
+\dots \,\,.
\label{eq:exactdalpha}
\end{multline}

Each time step in \eqref{eq:exactdalpha}
can be regarded as an independent
differential time evolution
so that,  from the initial condition
\eqref{eq:incondalpha}, 
we can make $\ve{\alpha}(t_{m-1})=0$
without any loss of generality.
Because of this,
$V^{-1}\left(\ve{\alpha}(t_{m-1})\right)=V^{-1}\left(0\right)=I$
and
$M^\top\left(\ve{\alpha}(t_{m-1})\right)=M^\top\left(0\right)=I$
and consequently
the first order term in
Eq. \eqref{eq:exactdalpha} gives
\begin{multline}
\ve{F}\left(t_{m-1},\boldsymbol{\alpha}(t_{m-1})\right)
\frac{\Delta t}{1!}\\
= V^{-1}\left(\ve{\alpha}(t_{m-1})\right)
M^\top\left(\ve{\alpha}(t_{m-1})\right) \boldsymbol{a}(t_{m-1})
\Delta t\\
=\boldsymbol{a}(t_{m-1})\Delta t.
\end{multline}
The second order term is
\begin{multline}
\dot{\ve{F}}\left(t_{m-1},\boldsymbol{\alpha}(t_{m-1})\right)
\frac{\Delta t^2}{2!}
=\bigg[\dot{V}^{-1}\left(\ve{\alpha}(t)\right)
M^\top\left(\ve{\alpha}(t)\right) \ve{a}(t)\\
+V^{-1}\left(\ve{\alpha}(t)\right)
\dot{M}^\top\left(\ve{\alpha}(t)\right) \ve{a}(t)\\
+V^{-1}\left(\ve{\alpha}(t)\right)
M^\top\left(\ve{\alpha}(t)\right) \dot{\ve{a}}(t)\bigg]
\frac{\Delta t^2}{2!}\\
=\bigg[\dot{V}^{-1}\left(\ve{\alpha}(t)\right)
\ve{a}(t)
+\dot{M}^\top\left(\ve{\alpha}(t)\right) \ve{a}(t)\\
+ \dot{\ve{a}}(t)\bigg]\frac{\Delta t^2}{2!}.
\label{eq:secondordterm}
\end{multline}
Using  \eqref{eq:vexact}, \eqref{eq:defI}
and the identity
\begin{multline}
\dot{V}^{-1}\left(\ve{\alpha}(t_{m-1})\right)\\
=-V^{-1}\left(\ve{\alpha}(t_{m-1})\right)
\dot{V}\left(\ve{\alpha}(t_{m-1})\right)
V^{-1}\left(\ve{\alpha}(t_{m-1})\right)\\
=-\dot{V}\left(\ve{\alpha}(t_{m-1})\right),
\end{multline}
the first element of \eqref{eq:secondordterm}
reduces to
\begin{multline}
\bigg(\dot{V}^{-1}\left(\ve{\alpha}(t)\right)
\ve{a}(t)\bigg)_i
=\sum_{k=1}^{n-1}\sum_{j=k+1}^{n}
c_{i,j,k}a_j(t)a_k(t).
\end{multline}
The second element of \eqref{eq:secondordterm}
vanishes, i.e.,
\begin{equation}
\bigg(\dot{M}^\top\left(\ve{\alpha}(t)\right) 
\ve{a}(t)\bigg)_i
=\sum_{k=1}^{n}\sum_{j=1}^{n}
c_{i,j,k}a_j(t)a_k(t) = 0,
\end{equation}
because $c_{i,j,k}a_j(t)a_k(t)$ is fully antisymmetric. 
Thus, the second order term is given by
\begin{multline}
\dot{F}_i\left(t_{m-1},\ve{\alpha}(t_{m-1})\right)
\frac{\Delta t^2}{2!}\\
=\Bigg[ \sum_{k=1}^{n-1}\sum_{j=k+1}^{n}
c_{i,j,k}a_j(t)a_k(t)+\dot{a}_i(t)\Bigg]\frac{\Delta t^2}{2!}.
\end{multline}
Gathering all the contributions
up to sencond order in $\Delta t$
yields
\begin{multline}
\big(d\ve{\alpha}_m\big)_i
=\Delta t\,  a_i(t_{m-1})
+ \frac{\Delta t^2}{2}\dot{a}_i(t_{m-1})\\
+\frac{\Delta t^2}{2}\sum_{k=1}^{n-1}\sum_{j=k+1}^{n}
c_{i,j,k}a_j(t)a_k(t) + O(\Delta t^3),
\label{eq:sndorderapprox}
\end{multline}
where $\big(d\ve{\alpha}_m\big)_i$ is the $i$-th
component of $d\ve{\alpha}_m$ with $i=1,2,\dots n$.
Equation \eqref{eq:alphadotexact} and
consequently \eqref{eq:sndorderapprox} take into account
the ordering of the Hamiltonian gates prescribed
by Eq. \eqref{eq:transham} hence enabling us
to avoid Trotterization provided that
we aim to errors of the order of $\Delta t^3$.

More accurate approximations of $d\ve{\alpha}(t_m)$ could 
iteratively
be built  up to
a certain  tolerated error of $\Delta t$
reducing the number of time steps and quantum gates.
However, 
this would imply a heavy
load of operations in the
classical stage of the computation
that could
undermine the effectiveness
of the algorithm.
Hence, it should be established
at what order of $\Delta t$
the costs of calculating $d\ve{\alpha}_m$
outweigh the gain in precision.

\section{Quantum circuit}
The fundamental stages of the quantum circuit are presented in
Fig. \ref{fig01}.
The Hamiltonian and $M(\boldsymbol{\alpha}(t))$
are considered to
have arbitrary dimension $n=2^{n_q-1}$
where $n_q$ is the number of required qbits
to perform the quantum simulation considering
that an extra control qbit is needed
for the quantum tomography of $\ket{\ve{\rho}}$.
The control qbit and the remaining ones
will be termed first and second registers,
respectively.

The first stage is the initial
state preparation.
In it, the goal is to set
the initial state to
the superposition
\begin{equation}
\left\vert \psi(0) \right\rangle
=\left\vert 0 \right\rangle\otimes
  \big[\left\vert u\right\rangle\big]
+\left\vert 1 \right\rangle\otimes
  \big[\left\vert \ve{\rho}(0)\right\rangle\big].
  \label{eq:initstate}
\end{equation}
The second part of the previous state
initializes the density matrix
coefficients to
\begin{equation}
\left\vert \ve{\rho}(0)\right\rangle 
= \sum_{i=1}^{n}\rho_i(0)\left\vert h_i\right\rangle
\end{equation}
and the first
is used as an ancillary uniform superposition
\begin{equation}
\left\vert u\right\rangle 
= 2^{-n/2}\sum_{i=1}^n\left\vert h_i\right\rangle,
\end{equation}
later on used to carry out the phase kickback.
Conveniently, 
the density matrix coefficients $\rho_i(0)$ are
arranged so that the
binary string of $i-1$ matches
the second register.

The first Hadamard gate ($H$-gate) puts the
control qbit
in the superposition 
\begin{equation}
\left\vert \psi_1\right\rangle
=\left(\left\vert 0 \right\rangle+\left\vert 1 \right\rangle\right)/\sqrt{2}
\otimes[\left\vert 0 \right\rangle^{\otimes (n_q-1)}].
\end{equation}
From this point on the part of the state
within square braces refers to the
second register.
The role of the controlled $U_H$ gate is to
set the initial condition for the normalized
components of the density matrix vector
$\left\vert \boldsymbol{\rho}(0)\right\rangle$.
This section of the circuit can be represented by
the unitary transformation
\begin{equation}
P_0\otimes [I]+P_1 \otimes [U_H]
\end{equation}
where $P_0=\left\vert 0 \right\rangle\left\langle 0\right\vert$
and $P_1=\left\vert 1 \right\rangle\left\langle 1\right\vert$
are the standard projectors.
The purpose of this gate is
to initialize the second register,
whose control qbit is $\left\vert 1 \right\rangle$,
to the normalized coefficients of
$\left\vert \boldsymbol{\rho}(0)\right\rangle$
living the other unmodified.
Given that the second register is initially
set to $\left\vert 0 \right\rangle^{\otimes (n_q-1)}$
by default, the first column vector of the $U_H$
must be ($\rho_1(0), \rho_2(0), \dots , \rho_n(0))$.
There are many variants of the
$U_H$ matrix that meet this requirement
and that of being unitary.
One option is the Householder matrix
given by
\begin{equation}
U_H =I
- \frac{\left(\boldsymbol{\rho}-\boldsymbol{u}_1\right)
\left(\boldsymbol{\rho}-\boldsymbol{u}_1\right)^\top}{
\left(1-\rho_1\right)},
\end{equation}
where $\boldsymbol{u}_1=(1,0,\dots,0)^\top$.
The singularity produced when
$\rho_1=1$ and $\rho_2=..=\rho_n=0$
can be avoided by choosing $\boldsymbol{u}_2$,...
or $\boldsymbol{u}_n$ instead of $\boldsymbol{u}_1$.
The following controlled $H$-gates
set the part of the second register
that is proportional to the
control qbit $\left\vert 0 \right\rangle$
to a uniform superposition.
This segment of the circuit
can be cast in the form of the transformation
\begin{multline}
X\otimes[I]
(P_0\otimes[I]+P_1\otimes[H^{\otimes (n_q-1)}]) 
X\otimes[I]\\
=P_0\otimes[H^{\otimes (n_q-1)}]+P_1\otimes[I].
\end{multline}
After collecting the gate transformations
and from some elementary algebra it follows that
the initial state is given by
\begin{multline}
\left\vert\psi(0)\right\rangle
=\left(P_0\otimes [H^{\otimes(n_q-1)}]+P_1 \otimes [I]\right)\\
\times
\left(P_0\otimes [I]+P_1 \otimes [U_H]\right)
\left\vert 0\right\rangle^{\otimes n_q}\\
=\frac{1}{\sqrt{2}}\left\vert 0 \right\rangle
\otimes\left[
H^{\otimes (n_q-1)}
\left\vert 0 \right\rangle^{\otimes (n_q-1)}\right]\\
+\frac{1}{\sqrt{2}}\left\vert 1 \right\rangle\otimes
\left[U_H\left\vert 0 \right\rangle^{\otimes (n_q-1)}\right]\\
=\frac{\left\vert 0 \right\rangle}{\sqrt{2}}
\otimes\left[
\sum_{i=1}^{n}\frac{1}{2^{n/2}}
\left\vert h_i \right\rangle 
\right]
+\frac{\left\vert 1 \right\rangle}{\sqrt{2}}\otimes
\bigg[\sum_{i=1}^{n}\rho_i(t)
\left\vert h_i \right\rangle 
\bigg]\\
=\frac{1}{\sqrt{2}}\left\vert 0 \right\rangle
\otimes\big[
\left\vert u \right\rangle 
\big]
+\frac{1}{\sqrt{2}}\left\vert 1 \right\rangle\otimes
\big[\left\vert \ve{\rho}(0) \right\rangle 
\big],
\end{multline}
which is consistent with \eqref{eq:initstate}.

The next portion of the circuit
is devoted to the actual time evolution
of $\left\vert \ve{\rho}(t)\right\rangle$.
The action of the controlled $M(\ve{\alpha}(t))$ gate
on the initial state $\left\vert \psi(0)\right\rangle$
can be expressed as
\begin{multline}
\left(P_0\otimes\left[I\right]
  +P_1\otimes\left[M(\ve{\alpha})\right]\right)
  \left\vert \psi(0) \right\rangle\\
  =\frac{\left\vert 0 \right\rangle}{\sqrt{2}}
\otimes\left[
\sum_{i=1}^{n}\frac{1}{2^{n/2}}\left\vert h_i \right\rangle 
\right]
+\frac{\left\vert 1 \right\rangle}{\sqrt{2}}
\otimes
\bigg[M(\ve{\alpha}(t))\left\vert \ve{\rho}(0) \right\rangle 
\bigg]\\
=\frac{\left\vert 0 \right\rangle}{\sqrt{2}}
\otimes\big[
\left\vert u \right\rangle 
\big]
+\frac{\left\vert 1 \right\rangle}{\sqrt{2}}\otimes
\big[\left\vert \ve{\rho}(t) \right\rangle 
\big].
\end{multline}
As it is shown in Fig. \ref{fig02},
the controlled $M(\boldsymbol{\alpha})$ gate
is broken down into small time steps
according to Eq. \eqref{eq:rhotimesteps}.
In agreement with Eqs. \eqref{eq:mkgate} and
\eqref{eq:mexact} each differential time step 
$M(d\ve{\alpha})$ is decomposed
into $n-1$ Hamiltonian gates of the form
\begin{equation}
M_k\left(d\alpha_k\right) 
= \exp\left(-\frac{i}{\hbar} H_kd\alpha_k\right)
\label{eq:hamgate1}
\end{equation}
where the Hamiltonian operator is
connected to the structure constants through 
\begin{equation}
H_k = -i C_k,
\label{eq:hamgate2}
\end{equation}
as can be seen in Fig. \ref{fig03}.
These gates
can be synthetized
for $SU(2^N)$ operations with arbitrary
number of qbits by means of
Cartan decomposition
\cite{KHANEJA200111,PhysRevA.69.010301,
Drury_2008,Dagli_2008},
and other alternative methods
\cite{PhysRevLett.114.090502,PhysRevLett.129.070501}.
Yet, the complexity and depth of the quantum
circuits increases
exponentially with the number of qbits
producing fast fidelity decays.
Fortunately, the structure constants
$C_k$ acquire properties that facilitate
the implementation of the Hamiltonian gates
if we restrict ourselves to
the algebra $\mathfrak{h}_n$ whose elements
are the Pauli strings
$h_k = (1/2)^{n/2}
\sigma_{k_1}\otimes\sigma_{k_2}\dots \otimes\sigma_{k_n}$.
The structure constants are
$n\times n$ matrices that can be expanded
in terms of the elements of the algebra
$\mathfrak{h}_n^{\otimes 2}$
$=\mathfrak{h}_n\otimes\mathfrak{h}_n$
$=\left\{h_1\otimes h_1, h_1\otimes h_2,
\dots , h_{n-1}\otimes h_n, h_n\otimes h_n
\right\}$
$=\{h_1^{(2)}, h_2^{(2)},\dots , h_{n^2}^{(2)}\}$ as
\begin{equation}
C_k=\sum_{i=1}^{n^2} \Tr\left[C_k h_i^{(2)}\right]h_i^{(2)}.
\end{equation}
Thereby the Hamiltonian gate becomes
\begin{multline}
\exp\left(-\frac{i}{\hbar} H_kd\alpha_k\right)\\
=
\exp\left(-\frac{1}{\hbar}
\sum_i
\Tr\left[C_k h_i^{(2)}\right]
h_i^{(2)}
d\alpha_k\right),
\end{multline}
where $i$ spans only the non-vanishing
coefficients $\Tr[C_k h_i^{(2)}]$.
Furthermore, in the Appendix \ref{ap:paulident} we prove that
in the previous expansion
the elements $h_i^{(2)}$ with non-vanishing coefficients
$\Tr[C_k h_i^{(2)}]$ commute with each other.
This is quite advantageous because it allows
to build the Hamiltonian gate \eqref{eq:hamgate1}
as the sequence of Pauli gates
\begin{multline}
\exp\left(-\frac{i}{\hbar} H_kd\alpha_k\right)\\
=\prod_{i}
\exp\left(-\frac{1}{\hbar}
\Tr\left[C_k h_i^{(2)}\right]
h_i^{(2)}
d\alpha_k\right),\label{eq:hamgate3}
\end{multline}
without regard to the ordering thereof,
and consequently
without the need of Trotterization. 
Additionally the Hamiltonian gates of
Pauli strings can be very efficiently simulated
using Clifford gates \cite{nielsen2002quantum} .

Up to this point (after the last $H$-gate),
the wave function takes the form
\begin{multline}
\left\vert \psi(t) \right\rangle 
=\frac{\left\vert 0 \right\rangle}{2}
\otimes\left[
\sum_{i=1}^{n}\left(\frac{1}{2^{n/2}}+\rho_i(t)\right)
\left\vert h_i \right\rangle 
\right]\\
+\frac{\left\vert 1 \right\rangle}{2}\otimes
\bigg[\sum_{i=1}^{n}\left(\frac{1}{2^{n/2}}-\rho_i(t)\right)
\left\vert h_i \right\rangle 
\bigg].
\end{multline}
Finally, the
density matrix coefficients 
$\rho_i(t)$ are computed as
\begin{equation}
\rho_i(t) = 2^{n/2}\left(p_{0,i}-p_{1,i}\right),
\end{equation}
where $p_{0,i}$ and $p_{1,i}$
are the probabilities corresponding to
$\left\vert 0\right\rangle\otimes [\left\vert h_i\right\rangle]$
and
$\left\vert 1\right\rangle\otimes [\left\vert h_i\right\rangle ]$,
respectively.
These are given by
\begin{eqnarray}
p_{0,i} &=& \left\vert\left\langle 0\right\vert\otimes
\left[\left\langle h_i \right\vert\right]\left\vert \psi(t) \right\rangle
\right\vert^2,\\
p_{1,i} &=& \left\vert\left\langle 1\right\vert\otimes
\left[\left\langle h_i \right\vert\right]\left\vert \psi(t) \right\rangle
\right\vert^2,
\end{eqnarray}
and are computed through the counts
from the circuit execution.

\begin{figure}
  \includegraphics[width=0.48\textwidth]{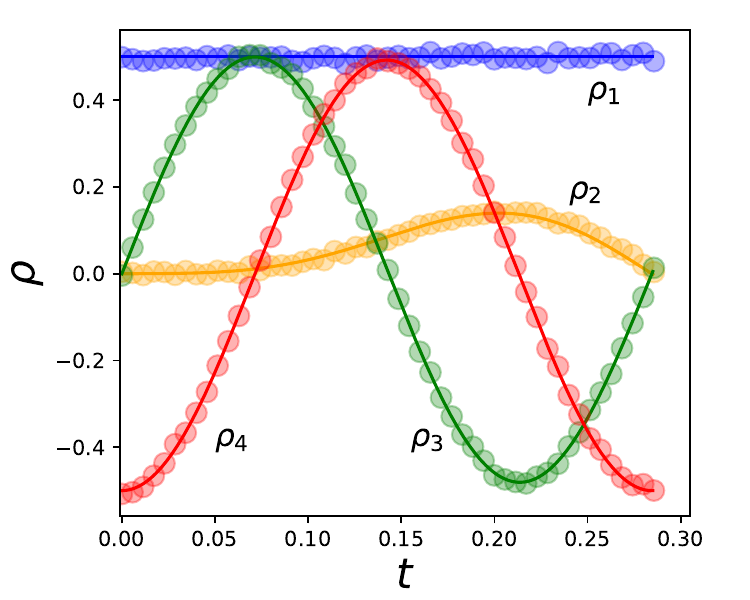}
  \caption{
    Density matrix coefficients as functions of time
    for the magnetic resonance Hamiltonian.
    The plots from the classical (continuous lines)
    and quantum (circles) computations
    are shown.
    }
    \label{fig04}
\end{figure}

\section{Example 1: one spin 1/2 particle
subject to a time varying magnetic field}
As a toy example, in this section
we examine
a spin $1/2$ particle
in a varying magnetic field.
The time
evolution of the density matrix elements
$\rho_1(t)$, $\rho_2(t)$, $\rho_3(t)$ and $\rho_4(t)$
were calculated by means of the classical and
quantum algorithms.
The Hamiltonian for this system is
\begin{multline}
H(t)=\omega_1 \cos(\omega t)\sigma_x/2\\
    -\omega_1 \cos(\omega t + \phi)\sigma_y/2
    -\omega_0 \sigma_z/2.
\end{multline}
Projecting it onto the base formed
by the Pauli matrices
by means of Eq. \eqref{eq:hamprojectionh}
we obtain
$a_1(t)=0$,
$a_2(t)=\omega_1 \cos(\omega t)/\sqrt{2}$,
$a_3(t)=-\omega_1 \cos(\omega t+\phi)/\sqrt{2}$
and
$a_4(t)=\omega_0/\sqrt{2}$.
If we assume that at $t=0$ the particle
is in the lowest energy level,
the initial condition for
the density matrix must be
\begin{equation}
\rho(0) = \frac{1}{2}\left(I-\sigma_z\right).
\end{equation}
The initial density matrix coefficients
are thus obtained from Eq. \eqref{eq:hamprojectionh}
giving
$\rho_1(0)=1/\sqrt{2}$,
$\rho_2(0)=0$,
$\rho_3(0)=0$ and
$\rho_4(0)=-1/\sqrt{2}$.
In this example we have set
the following Hamiltonian parameters:
$\omega=0.9$, $\omega_0=1$,
$\omega=22.0$ and
$\phi=\pi/2$.

In Fig. \ref{fig04} we present the density matrix
coefficients as a function of time calculated
through the classical (continuous lines)
and quantum (circles) algorithms.
The quantum algorithm was executed in the
IBM noisy quantum circuit (QASM) simulator.
Each dot in the plot was obtained
from a total of $16384$ shots. 
The figure shows that
the classical and quantum algorithms are
consistent.
\begin{figure}
  \includegraphics[width=0.45\textwidth]{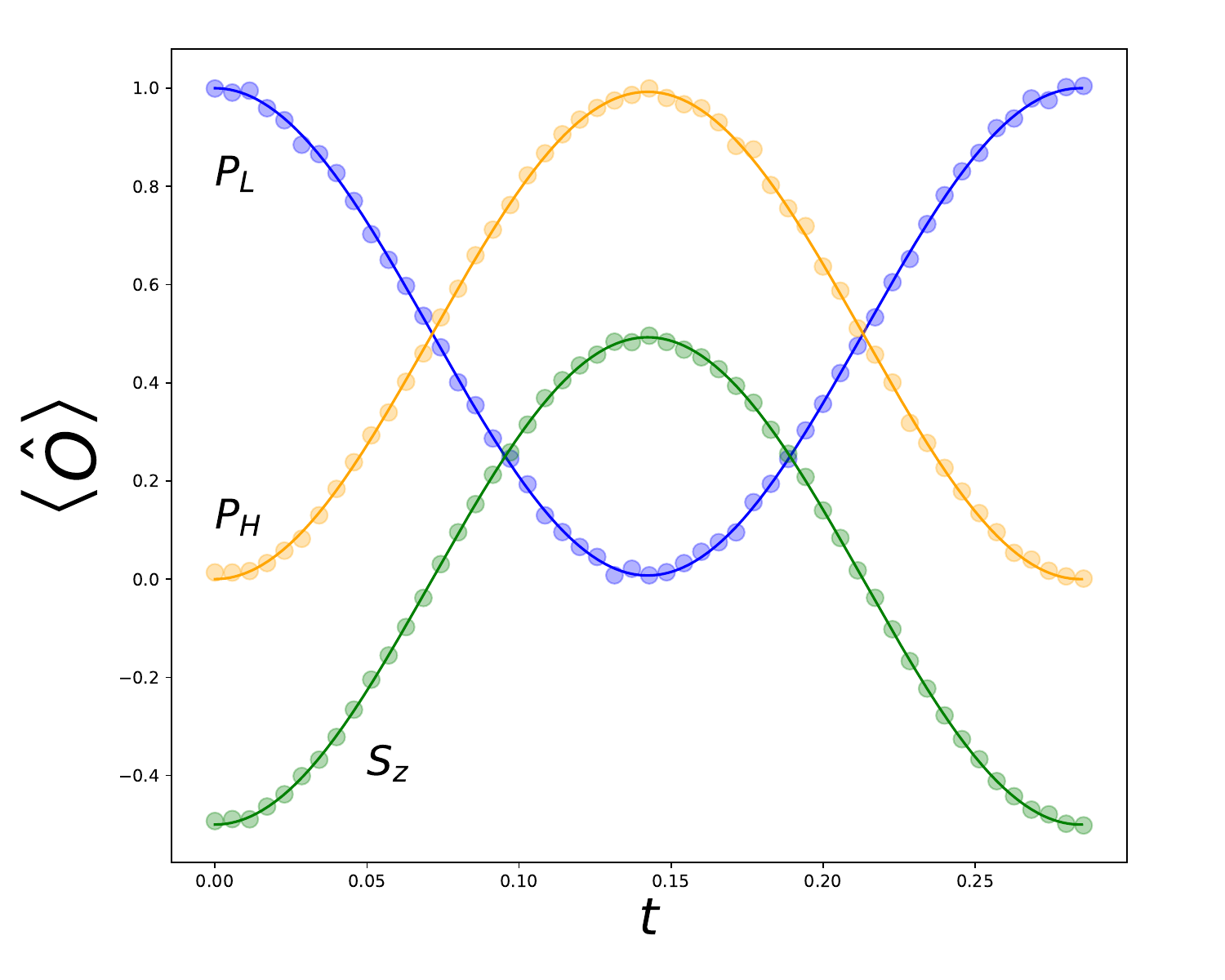}
  \caption{
    Expected values of the lowest energy level population,
    highest energy level population
    and spin projection along the $z$ axis.
    The classical and quantum computations
    are shown as solid lines and circles, respectively.
    }
    \label{fig05}
\end{figure}
\begin{figure*}
  \includegraphics[width=0.95\textwidth]{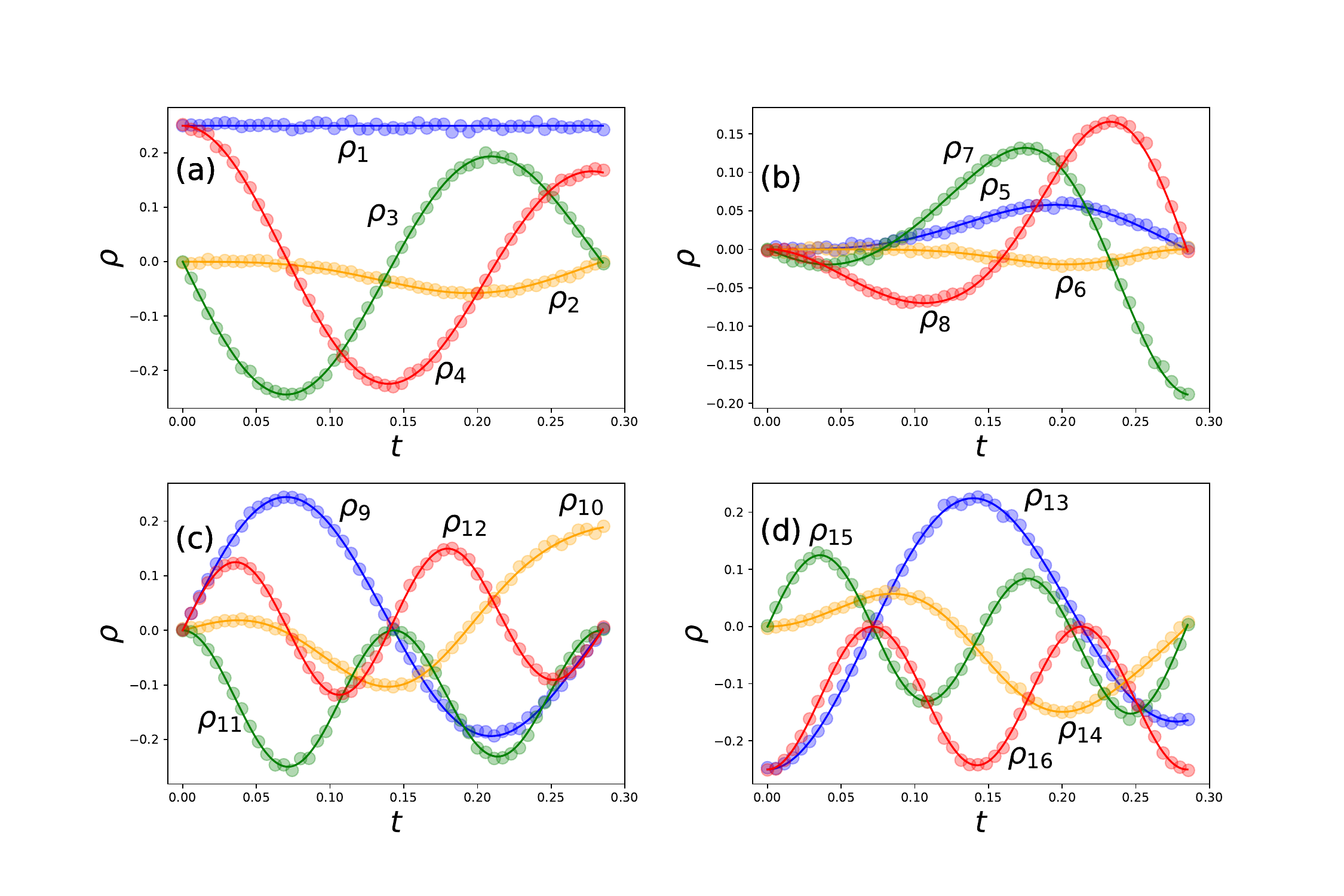}
  \caption{
    Density matrix coefficients as functions of time
    for the Hamiltonian corresponding to
    two spin 1/2 particles
    coupled through exchange interaction
    subject to a oscillating magnetic field.
    The classical and quantum computations
    are shown as solid lines and circles, respectively.
    }
    \label{fig06}
\end{figure*}
\begin{figure}
  \includegraphics[width=0.46\textwidth]{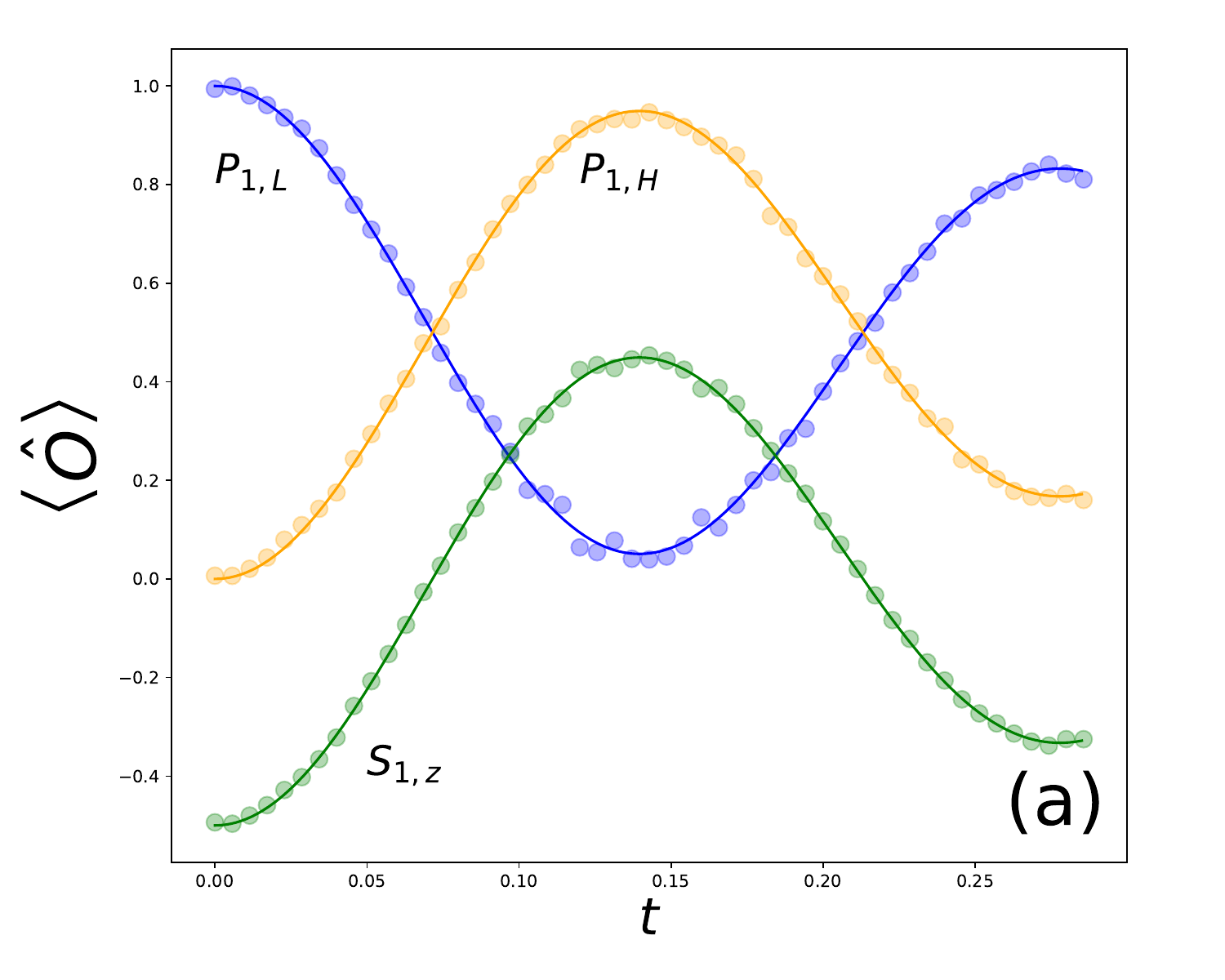}
  \includegraphics[width=0.46\textwidth]{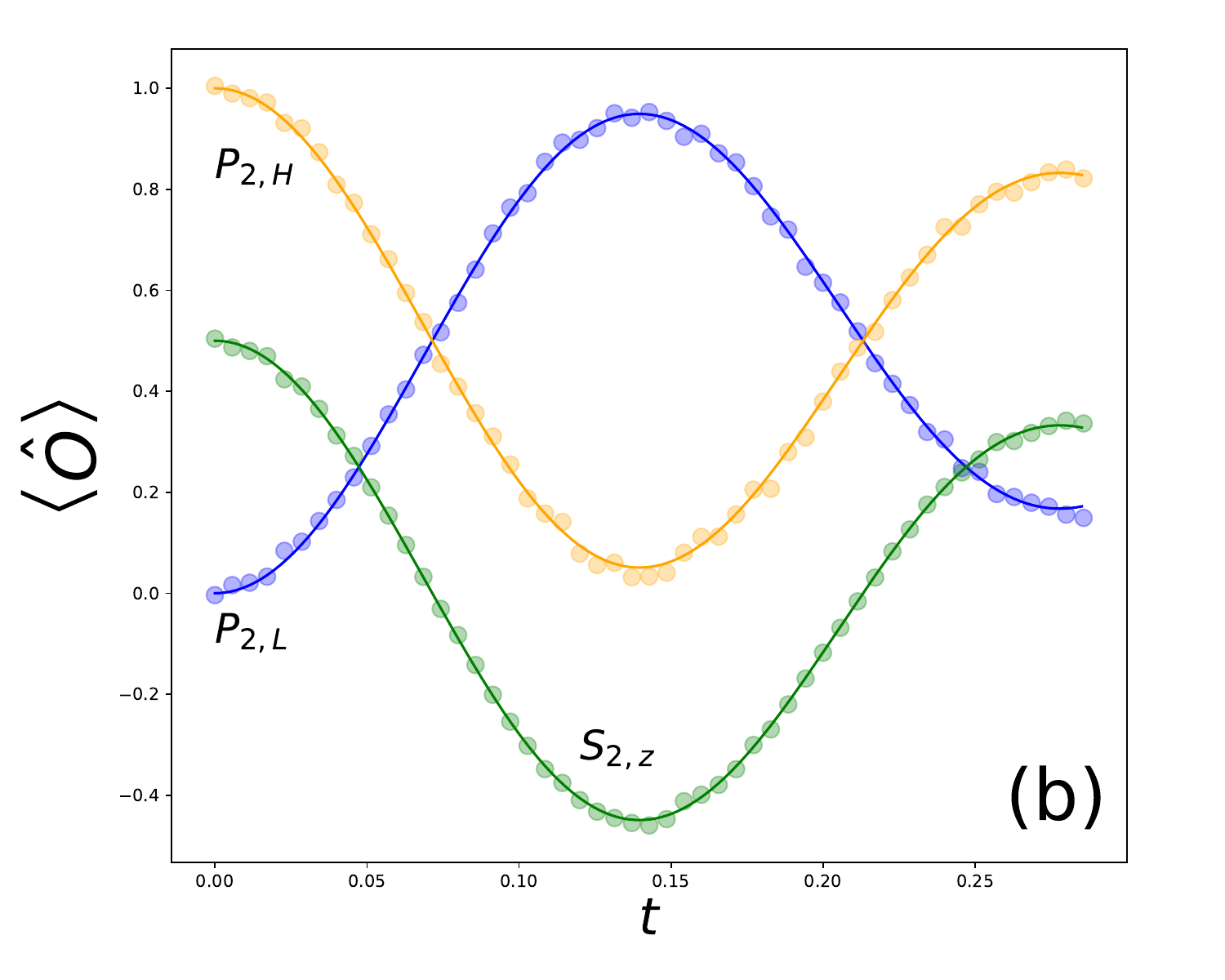}
  \caption{
    Expected values of the lowest energy level population,
    highest energy level population
    and spin projection along the $z$ axis for
    (a) the first and (b) second atoms.
    The classical and quantum computations
    are shown as solid lines and circles, respectively.
    }
    \label{fig07}
\end{figure}

In order to illustrate how to
calculate expected values
using the density matrix coefficients
and to further compare the
classical and quantum algorithms
Fig. \ref{fig05} exhibits the
expected values of the populations
of the lowest and highest energy levels
and the $z$ projection of the spin
given by
\begin{eqnarray}
P_{L}(t) &=& \mathrm{Tr}\left[
\left\vert 1 \right\rangle
\left\langle 1\right\vert \rho(t)\right], \\
P_{H}(t) &=& \mathrm{Tr}\left[\left\vert 2 \right\rangle
\left\langle 2\right\vert  \rho(t)\right], \\
S_{z}(t) &=& \mathrm{Tr}\left[
\frac{1}{2}\sigma_z  \rho(t)\right].
\end{eqnarray}
All the results in this section
where also confirmed by directly solving
the von Neumann equation \eqref{eq:vonneumanneq}
for the matrix elements of $\rho(t)$
and then projecting on to the
elements of $\mathfrak{h}_n$ using
Eq. \eqref{eq:rhoprojectionh}.

\section{Example 2: two spin 1/2 particles
coupled through exchange interaction
subject to a oscillating magnetic field.}

In order to test a larger system,
in this section we deal
with the time evolution
of two spin $1/2$ particles
coupled through exchange interaction
subject to a time-varying magnetic field.
The Hamiltonian for two electron spins
coupled through the exchange interaction
characterized by $A$
and subject to a varying magnetic
field is given by
\begin{multline}
H=\omega_1\cos(\omega t)
 \left(S_{x,1}+S_{x,2}\right)\\
 +\omega_1\cos(\omega t+\phi)
 \left(S_{y,1}+S_{y,2}\right)
 +\omega_0\left(S_{z,1}+S_{z,2}\right)\\
 +A \ve{S}_1\cdot\ve{S}_2,
\end{multline}
where $\ve{S}_1$ and $\ve{S}_2$ are the
spin operators for the first and second
particles, respectively.
Projecting onto the base formed by
the Pauli strings 
the non vanishing coefficients of the Hamiltonian are
$a_2(t) = a_5(t) = \omega_1 \cos(\omega t)$,
$a_3(t) = a_9(t) =-\omega_1 \cos(\omega t+\phi)$,
$a_4(t) = a_{13}(t) = -\omega_0$ and
$a_6(t) = a_{11}(t) = a_{16}(t)= A/2$.

The initial condition, that corresponds
to both electrons being in the lowest
energy level, is given by
\begin{equation}
\rho(0) = \frac{1}{2}\left(I -\sigma_z\right)
\otimes \frac{1}{2}\left(I -\sigma_z\right).
\end{equation}
The non vanishing coefficients corresponding
to this condition are
$\rho_1(0) = \rho_4(0) = -\rho_{13}(0) =  -\rho_{16}(0)=1/2$.
In this example the Hamiltonian parameters
are
$\omega=0.9$, $\omega_0=1$,
$\omega=22.0$,
$\phi=\pi/2$
and $A=3.0$.

In Fig. \ref{fig06} we observe the $16$ density matrix coefficients
as a function of time. Both classical and quantum computations
are consistent.
To further compare the classical and quantum algorithms
Fig. \ref{fig07} presents the expected values of
the population corresponding to
the lowest and highest energy levels of both atoms
$P_{1,L}(t)$, $P_{1,H}(t)$, $P_{2,L}(t)$, $P_{2,H}(t)$
where
\begin{eqnarray}
P_{1,L}(t) &=& \mathrm{Tr}\left[
\left\vert 1 \right\rangle
\left\langle 1\right\vert \otimes I \rho(t)\right], \\
P_{1,H}(t) &=& \mathrm{Tr}\left[\left\vert 2 \right\rangle
\left\langle 2\right\vert \otimes I \rho(t)\right], \\
P_{2,L}(t) &=& \mathrm{Tr}\left[
I\otimes\left\vert 1 \right\rangle
\left\langle 1\right\vert \rho(t)\right], \\
P_{2,H}(t) &=& \mathrm{Tr}\left[
I\otimes\left\vert 2 \right\rangle
\left\langle 2\right\vert \rho(t)\right].
\end{eqnarray}
This figure also shows the expected values
of the spin $z$ component $S_z$
\begin{eqnarray}
S_{1,z}(t) &=& \mathrm{Tr}\left[
\frac{1}{2}\sigma_z \otimes I \rho(t)\right], \\
S_{2,z}(t) &=& \mathrm{Tr}\left[I \otimes
\frac{1}{2}\sigma_z  \rho(t)\right],
\end{eqnarray}
where $I$ is the $2\times 2$ identity matrix.

As in the previous example, the coefficients
of the density matrix where also confirmed
by solving Eq. \eqref{eq:vonneumanneq}
and then projecting onto the elements of
$\mathfrak{h}_n$ through Eq. \eqref{eq:rhoprojection}.

\section{Conclusions}

In the present paper, a general framework to linearize the von-Neumann
equation was developed. It mainly relies on the projection of
the density matrix and the Hamiltonian on an operator base
formed by the elements of a Lie algebra.
For the particular case of $\mathfrak{q}_n$ the von-Neumann equation
is mapped to the conventional Shr\"odinger-like equation \eqref{eq:stackedvonneumann},
where the state vector $\left\vert\ve{\rho}(t)\right\rangle$ corresponds
to the column stacked matrix elements of the density matrix $\rho (t)$
and the Hamiltonian to the superoperator
$\mathcal{H}=I\otimes H(t)-H^\top(t)\otimes I$.
It is shown that this is but one of the multiple ways
of linearizing the von-Neumann equation. Other
versions can be obtained by projecting
onto different algebras.
In the case of the Hermitian algebras $\mathfrak{h}_n$
the linearization yields
a density matrix vector $\left\vert\ve{\rho}(t)\right\rangle$ with purely
real entries which highly simplifies the
quantum tomography of the final state vector.
Moreover, it was proven that although the
Hamiltonian superoperator has dimension $n\times n$, only
$n$ generators are needed to build the time evolution
operator considerably reducing the number of necessary
quantum gates.
Due to the unique properties of the Pauli strings
the Hamiltonian gates can be implemented
as a sequence of at most $n$ commuting Pauli gates.
This presents two advantages: first,
Pauli gates are easy to create
in terms of Clifford gates and, second,
no Trotterization is needed to mitigate errors
considerably reducing the circuit depth. 

All these notions were used to implement a quantum algorithm that
solves the von-Neumann equation.
The quantum algorithm was tested against
the classical solution
of the von Neumann equation for two toy Hamiltonians
using the QASM simulator of IBMQ \cite{ibm}
giving identical results.
Even though
these algorithms were devised for the algebra whose
elements are the Pauli strings, they can readily
be adapted for any other Hermitian algebra.

We have seen in Eqs. \eqref{eq:hamgate1} and \eqref{eq:hamgate2}
that the structure constants
generate the time evolution of the density matrix
through a series of unitary quantum gates.
This framework thus gives us the flexibility to
engineer the structure constants
and consequently the required
quantum gates
by choosing the elements of the algebra.

This same scheme can also be applied to
the analysis of the time evolution of open-quantum systems
through the linearization of the
Lindblad-Von Neumann master equation.
This problem will be addressed elsewhere.

\begin{acknowledgments}
I thank A. Vega, A. Mart\'inez and S. Noyola
for valuable input
and inspiring discussions.
This work was financially supported
by Departamento de Ciencias B\'asicas 
UAM-A grant number 2232218.
I am indebted to IFUNAM for their hospitality. 
I acknowledge the use of IBM Quantum services for this work.
The views expressed are those of the author, and do not
reflect the official policy or position of IBM or the IBM 
Quantum team.
\end{acknowledgments}

\appendix

\section{Useful identities for the $\mathfrak{q}_n\otimes \mathfrak{q}_n$}
\label{ap:Qiden}
In this appendix we prove that the Kronecker product
of two matrices $B^\top \otimes A$ expanded in terms
of the elements $Q_{i,j}$ of $\mathfrak{q}_n\otimes \mathfrak{q}_n$
can be related to the expansion in terms of $q_k\otimes q_l$
through
\begin{equation}
\Tr\left[Q_{i,j}^\dagger B^\top \otimes A\right]
=\Tr\left[q_i^\dagger A q_j B\right].
\label{eq:Qqqrelap}
\end{equation}
Expanding $A$ and $B$ in the right-hand side
of \eqref{eq:Qqqrelap} we find that
\begin{multline}
\Tr\left[q_i^\dagger A q_j B\right]
=\sum_{a,b}\Tr\left[q_i^\dagger q_a^\dagger q_j q_b\right]
\Tr\left[q_a A\right]\Tr\left[q_b^\dagger B\right]\\
=\sum_{a,b}\Tr\left[q_i^\dagger  q_a^\dagger q_j q_b\right]
\Tr\left[q_a A\right]\Tr\left[q_b B^\top\right]\\
=\sum_{a,b}\Tr\left[q_i^\dagger q_a^\dagger q_j q_b\right]
\Tr\left[q_b\otimes q_a B^\top \otimes A\right].
\label{eq:Qqqrelap2}
\end{multline}
In the last step we have used
$\Tr\left[q_b B^\top\right]=\Tr\left[q_b^\dagger B\right]$.
Substituting the expansion of $q_b^\dagger\otimes q_a^\dagger$
in terms of the  $Q_{r,s}$ as
\begin{equation}
q_b\otimes q_a
= \sum_{k,l}\Tr\left[Q_{k,l}^\dagger q_b\otimes q_a\right]Q_{k,l}
\label{eq:kraus1}
\end{equation}
into Eq. \eqref{eq:Qqqrelap2} we obtain
\begin{multline}
\Tr\left[q_i^\dagger A q_j B\right]
=\sum_{k,l,a,b}\Tr\left[q_i^\dagger q_a^\dagger q_j q_b\right]
\Tr\left[Q_{k,l}^\dagger q_b\otimes q_a\right]\\
\times\Tr\left[Q_{k,l} B^\top \otimes A\right].
\label{eq:Qqqrelap3}
\end{multline}
Only if $a$ and $b$ are given by \eqref{eq:Qqqrel2} and \eqref{eq:Qqqrel3},
as stated by \eqref{eq:Qqqrel1}, 
the term $\Tr[Q_{k,l}^\dagger q_b\otimes q_a]=1$,
otherwise  $\Tr[Q_{k,l}^\dagger q_b\otimes q_a]=0$. Hence,
the only non vanishing term from the sum in
\eqref{eq:Qqqrelap3} is
\begin{multline}
\Tr\left[q_i^\dagger q_a^\dagger q_j q_b\right]
=\sum_p\delta_{p,i_2}\delta_{i_1,a_2}\delta_{a_1,j_1}
\delta_{j_2,b_1}\delta_{b_2,p}\\
=\delta_{i_1,a_2}\delta_{a_1,j_1}
\delta_{j_2,b_1}\delta_{b_2,i_2}.
\end{multline}
Plugging \eqref{eq:i1} and \eqref{eq:i2} in to
the previous equation, one gets
\begin{multline}
\Tr\left[q_i^\dagger q_a^\dagger q_j q_b\right]
=\sum_p\delta_{p,i_2}\delta_{i_1,a_2}\delta_{a_1,j_1}
\delta_{j_2,b_1}\delta_{b_2,p}\\
=\delta_{i_1,a_2}\delta_{a_1,j_1}
\delta_{j_2,b_1}\delta_{b_2,i_2}\\
=\delta_{(i-1)\bmod N +1, [(a-1)/N]+1}\\
\times\delta_{(a-1)\bmod N +1, (j-1)\bmod N +1}\\
\times\delta_{[(j-1)/N]+1,(b-1)\bmod N +1 }\\
\times\delta_{[(b-1)/N]+1, [(i-1)/N]+1}.
\end{multline}
Substituting  the explicit expressions
for $a$ and $b$ from \eqref{eq:Qqqrel2} and \eqref{eq:Qqqrel3}
in the equation above yields
\begin{multline}
\Tr\left[q_i^\dagger q_a^\dagger q_j q_b\right]
=\delta_{(i-1)\bmod N +1, (k-1)\bmod N+1}\\
\times\delta_{(l-1)\bmod N +1, (j-1)\bmod N +1}\\
\times\delta_{[(j-1)/N]+1,[(l-1)/ N] +1 }\\
\times\delta_{[(k-1)/N]+1, [(i-1)/N]+1}=\delta_{i,k}\delta_{j,l}.
\end{multline}
Introducing this result
in \eqref{eq:Qqqrelap3}
we finally obtain \eqref{eq:Qqqrelap}.

\section{Useful properties of the Pauli strings algebra}
\label{ap:paulident}

In this appendix we derive useful properties
for the Pauli strings algebra 
mainly to obtain recurrence relations
that make the classical stage of the
algorithm more efficient.
Pauli strings have the form
$(1/2)^{n/2}
\sigma_{k_1}\otimes\sigma_{k_2}\dots \otimes\sigma_{k_n}$
where $\sigma_0$ is the $2\times 2$ identity matrix,
$\sigma_i$ ($i=1,2,3$) are proportional to 
the standard Pauli matrices
\begin{equation}
h_i^1=\frac{1}{\sqrt{2}}\sigma_i,
\end{equation}
and the factor $1/\sqrt{2}$ is used to normalize
the base to unity.
Here we have adopted a notation where the
superindex $n$ of the element $h_i^n$
indicates the number of qbits spanned by the algebra.
The elements of algebras of higher dimension
can be computed by sequentially taking the
Kronecker product of the elements of lower
dimensional algebras.
For example, the elements of the algebra
that spans two qbits may be obtained from
the Kronecker product of the elements
of the algebra corresponding to one qbit as
\begin{equation}
h_i^2=\frac{1}{\sqrt{2}}\sigma_{i_1}\otimes \frac{1}{\sqrt{2}}\sigma_{i_2}
=h_{i_1}^1\otimes h_{i_2}^1,
\end{equation}
where $i=4(i_1-1)+i_2$, $i_1, i_2=1,2,3,4$ and $i=1,2,\dots, 16$.
More generally, the algebra of $n+m$ qbits
can be generated from the 
Kronecker product of the algebras
of $n$ and $m$ qbits, i.e.,
\begin{equation}
h_i^{n+m}
= h_{i_1}^{n}\otimes h_{i_2}^m,
\label{eq:hnm}
\end{equation}
where
\begin{eqnarray}
i &=& 2^{2m}(i_1-1)+i_2,\\
i &=& 1,2,\dots, 2^{2(n+m)},\\
i_1 &=& 1,2,\dots, 2^{2n},\\
i_2 &=& 1,2,\dots, 2^{2m}.
\end{eqnarray}
In particular, a base corresponding to
an arbitrary number of qbits can be constructed
by recursively applying
\begin{equation}
h_i^{n+1}
= h_{i_1}^{n}\otimes h_{i_2}^1,
\end{equation}
where
$i = 4(i_1-1)+i_2$,
$i = 1,2,\dots, 2^{2(n+1)}$,
$i_1 = 1,2,\dots, 2^{2n}$,
$i_2 = 1,2,3,4$.

In this notation, the commutation of two elements of the
algebra is given in terms of the structure constants as
\begin{equation}
\left[h_i^n, h_j^n\right]=i\hbar \sum_{k=1}^{2^{2n}}c_{i,j,k}^nh_k^n .
\end{equation}
Using the orthonormailty of this base,
the explicit form of the structure constant
is given as
\begin{equation}
c_{i,j,k}^n
=\frac{1}{i\hbar}\Tr\left[\left[h_i^n, h_j^n\right]h_k^n\right].
\label{eq:cstrucons1}
\end{equation}
The structure constants $b_{i,j,k}^n$
ensued from taking the anticommutator as the Lie braket
allow to express the anticommutator as
\begin{equation}\left\{h_i^n, h_j^n\right\}
= \sum_{k=1}^{2^{2n}}b_{i,j,k}^nh_k^n .
\end{equation}
Explicitly,
\begin{equation}
b_{i,j,k}^n=\Tr\left[\left\{h_i^n, h_j^n\right\}h_k^n\right].
\label{eq:bstrucons1}
\end{equation}
Although these do not play any role in the
construction of the Hamiltonian gates,
they  will be very useful as auxiliary parameters
in determining recurrence relations
for $c_{i,j,k}^n$.

From the general commutator relations
\begin{multline}
\left[A_1\otimes A_2, D_1\otimes D_2\right]
=\frac{1}{2}\left[A_1, A_2\right]
\otimes\left\{D_1, D_2\right\}\\
+\frac{1}{2}\left\{A_1, A_2\right\}
\otimes\left[D_1,D_2\right],
\label{eq:generalcommu1}
\end{multline}
\begin{multline}
\left\{A_1\otimes A_2, D_1\otimes D_2\right\}
=\frac{1}{2}\left[A_1, A_2\right]
\otimes\left[D_1, D_2\right]\\
+\frac{1}{2}\left\{A_1, A_2\right\}
\otimes\left\{D_1,D_2\right\},
\label{eq:generalcommu2}
\end{multline}
if follows immediately that
the commutator and anticommutator 
of an algebra of dimension $2^{2(n+m)}$
can be expressed in terms of the
commutator and anticommutator of algebras
of smaller dimensions $2^{2n}$ and $2^{2m}$ as
\begin{multline}
\left[h_{i}^{n+m},h_{j}^{n+m}\right]
=\left[h_{i_1}^n\otimes h_{i_n}^m,
h_{j_1}^n\otimes h_{j_2}^n\right]\\
=\frac{1}{2}\left[h_{i_1}^n, h_{j_1}^m\right]
\otimes\left\{h_{i_2}^m,h_{j_2}^m\right\}\\
+\frac{1}{2}\left\{h_{i_1}^n,h_{j_1}^n\right\}
\otimes\left[h_{i_2}^m,h_{j_2}^m\right],\label{eq:kroncommu1}
\end{multline}
\begin{multline}
\left\{h_{i}^{n},h_{j}^{m}\right\}
=\left\{h_{i_1}^n\otimes h_{i_2}^m,
h_{j_1}^n\otimes h_{j_2}^n\right\}\\
= \frac{1}{2}\left[h_{i_1}^n,h_{j_1}^m\right]
\otimes\left[h_{i_2}^m,h_{j_2}^m\right]\\
+\frac{1}{2}\left\{h_{i_1}^n,h_{j_1}^n\right\}
\otimes\left\{h_{i_2}^m,h_{j_2}^m\right\}.\label{eq:kroncommu2}
\end{multline}

Now we move on to how to workout recurrence
relations for the structure constants.
Multiplying both sides of
\eqref{eq:kroncommu1} and \eqref{eq:kroncommu2}
by $h_k^{n+m} = h_{k_1}^n\otimes h_{k_2}^m$
and using the definitions \eqref{eq:cstrucons1}
and \eqref{eq:bstrucons1}
we obtain the following recurrence
relations
\begin{eqnarray}
c_{i,j,k}^{n+m}
= \frac{1}{2}\left(b_{i_1,j_1,k_1}^n c_{i_2,j_2,k_2}^m
+ c_{i_1,j_1,k_1}^n b_{i_2,j_2,k_2}^m
\right),\\
b_{i,j,k}^{n+1}
= \frac{1}{2}\left(b_{i_1,j_1,k_1}^n b_{i_2,j_2,k_2}^m
- c_{i_1,j_1,k_1}^n c_{i_2,j_2,k_2}^m
\right),
\end{eqnarray}
or more succintly
\begin{eqnarray}
C_{k}^{n+m}
= \frac{1}{2}\left(C_{k_1}^n \otimes B_{k_2}^m
+B_{k_1}^n\otimes C_{k_2}^m
\right),\label{eq:crecursive1}\\
B_{k}^{n+m}
= \frac{1}{2}\left(B_{k_1}^n \otimes B_{k_2}^m
-C_{k_1}^n \otimes C_{k_2}^m
\right),\label{eq:brecursive1}
\end{eqnarray}
where $(C_k^n)_{i,j}= c_{i,j,k}^n$ and $(B_k^n)_{i,j}= b_{i,j,k}^n$.
The use of these recurrence relations
to obtain structure constants is far more
efficient than the direct application of
\eqref{eq:cstrucons1} and \eqref{eq:bstrucons1}
because it avoids the tracing and all the matrix
multiplications required by \eqref{eq:cstrucons1}
and \eqref{eq:bstrucons1}.
For instance, one could initially calculate
the structure constants for $n=1$ (one qbit) through
\eqref{eq:cstrucons1} and \eqref{eq:bstrucons1},
and then, by setting $m=1$ in \eqref{eq:crecursive1}
and \eqref{eq:brecursive1} it
is possible to recursively compute
the commutors and anticommutors of
progressively larger algebras.

In a similar fashion we are able compute
recurrence relations for the coefficients
$\Tr[C_kh_i^{(2)}]=\Tr[C_k^n h_i^{2n}]$
of the structure constants
required by the Hamitlonian gate
\eqref{eq:hamgate3}.
To do so we substitute
\eqref{eq:brecursive1} with $m=1$
and  $h_i^{2n}=h_{i_1}^{n}\otimes h_{i_2}^{n}$
into $\Tr[C_k^{n+1} h_i^{2(n+1)}]$
obtaining the recurrence relation
\begin{multline}
\Tr\left[C_k^{n+1} h_i^{2(n+1)}\right]
=\frac{1}{2}\Tr\left[B_{k_1}^{n}h_{i_1}^{2n}\right]
\Tr\left[C_{k_2}^{1}h_{i_1}^{2}\right]\\
+\frac{1}{2}\Tr\left[C_{k_1}^{n}h_{i_1}^{2n}\right]
\Tr\left[B_{k_1}^{1}h_{i_2}^{2}\right].
\label{eq:ch1}
\end{multline}
Similarly $\Tr[B_k^{n+1} h_i^{2n}]$,
also required by the expression above,
can be computed as
\begin{multline}
\Tr\left[B_k^{n+1} h_i^{2(n+1)}\right]
=\frac{1}{2}\Tr\left[B_{k_1}^{n}h_{i_1}^{2n}\right]
\Tr\left[B_{k_2}^{1}h_{i_1}^{2}\right]\\
-\frac{1}{2}\Tr\left[C_{k_1}^{n}h_{i_1}^{2n}\right]
\Tr\left[C_{k_1}^{1}h_{i_2}^{2}\right].
\label{eq:bh1}
\end{multline}

It only remains to prove that
in the expansion of the Hamiltonian gate
\eqref{eq:hamgate3}
the elements $h_i^{2n}$ with non-vanishing coefficients
$\Tr[C_k^n h_i^{2n}]$ commute with each other.
This can be easily proven by showing
that the commutator of the projections of $C_k^n$ over two
generic elements 
$h_i^{2n}$ and $h_j^{2n}$ vanishes, namely
\begin{equation}
\bigg[\Tr\left[C_k^n h_i^{2n}\right]h_i^{2n},
\Tr\left[C_k^n h_j^{2n}\right]h_j^{2n}\bigg] =0.
\label{eq:commu0}
\end{equation}
By using Eq. \eqref{eq:cstrucons1}
Eq. \eqref{eq:commu0}
takes the form
\begin{multline}
\bigg[\Tr\left[C_k^n h_i^{2n}\right]h_i^{2n} \,,
\Tr\left[C_k^n h_j^{2n}\right]h_j^{2n}\bigg]\\
=i\hbar \Tr\left[C_k^n h_i^{2n}\right]
\Tr\left[C_k^n h_j^{2n}\right]
\sum_{m=1}^{2^{4n}}c_{i,j,m}^{2n}h_m^{2n}.
\end{multline}
Then, multiplying both sides by $h_l^{2n}$ and tracing,
we have
\begin{equation}
\Tr\left[C_k^n h_i^{2n}\right]
\Tr\left[C_k^n h_j^{2n}\right]
c_{i,j,l}^{2n}=0.\label{eq:projckh1}
\end{equation}
The values of a total of
$\overbrace{2^{2n}}^k\times \overbrace{2^{4n}}^i
\times \overbrace{2^{4n}}^j\times \overbrace{2^{4n}}^l=2^{14n}$
terms are needed to
prove that.
Even for a small number of qbits this is a
challenging task.
However, \eqref{eq:projckh1} can be recast
in the form of a recurrence relation by
substituting in it \eqref{eq:crecursive1},
\eqref{eq:ch1} and \eqref{eq:bh1} for which we
find
\begin{widetext}
\begin{multline}
\Tr\left[C_k^{n+1} h_i^{2(n+1)}\right]
\Tr\left[C_k^{n+1} h_j^{2(n+1)}\right]
c_{i,j,l}^{2(n+1)}
=\frac{1}{8}\Tr\left[C_{k_1}^n h_{i_1}^{2n}\right]
 \Tr\left[C_{k_1}^n h_{j_1}^{2n}\right] c_{i_1,j_1,l_1}^{2n}
 \Tr\left[B_{k_2}^1 h_{i_2}^{2}\right]
 \Tr\left[B_{k_2}^1h_{j_2}^{2}\right] b_{i_2,j_2,l_2}^{2}\\
+\frac{1}{8}\Tr\left[B_{k_1}^n h_{i_1}^{2n}\right]
 \Tr\left[B_{k_1}^n h_{j_1}^{2n}\right] c_{i_1,j_1,l_1}^{2n}
 \Tr\left[C_{k_2}^1 h_{i_2}^{2}\right]
 \Tr\left[C_{k_2}^1h_{j_2}^{2}\right] b_{i_2,j_2,l_2}^{2}\\
+\frac{1}{8}\Tr\left[C_{k_1}^n h_{i_1}^{2n}\right]
 \Tr\left[B_{k_1}^n h_{j_1}^{2n}\right] b_{i_1,j_1,l_1}^{2n}
 \Tr\left[B_{k_2}^1 h_{i_2}^{2}\right]
 \Tr\left[C_{k_2}^1h_{j_2}^{2}\right] c_{i_2,j_2,l_2}^{2}\\
+\frac{1}{8}\Tr\left[B_{k_1}^n h_{i_1}^{2n}\right]
 \Tr\left[C_{k_1}^n h_{j_1}^{2n}\right] b_{i_1,j_1,l_1}^{2n}
 \Tr\left[C_{k_2}^1 h_{i_2}^{2}\right]
 \Tr\left[B_{k_2}^1h_{j_2}^{2}\right] c_{i_2,j_2,l_2}^{2}\\
+\frac{1}{8}\Tr\left[B_{k_1}^n h_{i_1}^{2n}\right]
 \Tr\left[B_{k_1}^n h_{j_1}^{2n}\right] b_{i_1,j_1,l_1}^{2n}
 \Tr\left[C_{k_2}^1 h_{i_2}^{2}\right]
 \Tr\left[C_{k_2}^1h_{j_2}^{2}\right] c_{i_2,j_2,l_2}^{2}\\
+\frac{1}{8}\Tr\left[C_{k_1}^n h_{i_1}^{2n}\right]
 \Tr\left[C_{k_1}^n h_{j_1}^{2n}\right] b_{i_1,j_1,l_1}^{2n}
 \Tr\left[B_{k_2}^1 h_{i_2}^{2}\right]
 \Tr\left[B_{k_2}^1h_{j_2}^{2}\right] c_{i_2,j_2,l_2}^{2}\\
+\frac{1}{8}\Tr\left[B_{k_1}^n h_{i_1}^{2n}\right]
 \Tr\left[C_{k_1}^n h_{j_1}^{2n}\right] c_{i_1,j_1,l_1}^{2n}
 \Tr\left[C_{k_2}^1 h_{i_2}^{2}\right]
 \Tr\left[B_{k_2}^1h_{j_2}^{2}\right] b_{i_2,j_2,l_2}^{2}\\
+\frac{1}{8}\Tr\left[C_{k_1}^n h_{i_1}^{2n}\right]
 \Tr\left[B_{k_1}^n h_{j_1}^{2n}\right] c_{i_1,j_1,l_1}^{2n}
 \Tr\left[B_{k_2}^1 h_{i_2}^{2}\right]
 \Tr\left[C_{k_2}^1h_{j_2}^{2}\right] b_{i_2,j_2,l_2}^{2}.
 \label{eq:chchc1}
\end{multline}
The last four lines of the previous equation
vanish because, as can be demonstrated
from the direct computation of the traces and
structure constants,
\begin{multline}
\Tr\left[C_{k_2}^1 h_{i_2}^{2}\right]
\Tr\left[C_{k_2}^1h_{j_2}^{2}\right] c_{i_2,j_2,l_2}^{2}
=\Tr\left[B_{k_2}^1 h_{i_2}^{2}\right]
\Tr\left[B_{k_2}^1h_{j_2}^{2}\right] c_{i_2,j_2,l_2}^{2}\\
=\Tr\left[C_{k_2}^1 h_{i_2}^{2}\right]
\Tr\left[B_{k_2}^1h_{j_2}^{2}\right] b_{i_2,j_2,l_2}^{2}
=\Tr\left[B_{k_2}^1 h_{i_2}^{2}\right]
\Tr\left[C_{k_2}^1h_{j_2}^{2}\right] b_{i_2,j_2,l_2}^{2}=0.
\label{eq:recvanish}
\end{multline}
In this way, Eq. \eqref{eq:chchc1} simplifies into
\begin{multline}
\Tr\left[C_k^{n+1} h_i^{2(n+1)}\right]
\Tr\left[C_k^{n+1} h_j^{2(n+1)}\right]
c_{i,j,l}^{2(n+1)}
=\frac{1}{8}\Tr\left[C_{k_1}^n h_{i_1}^{2n}\right]
 \Tr\left[C_{k_1}^n h_{j_1}^{2n}\right] c_{i_1,j_1,l_1}^{2n}
 \Tr\left[B_{k_2}^1 h_{i_2}^{2}\right]
 \Tr\left[B_{k_2}^1h_{j_2}^{2}\right] b_{i_2,j_2,l_2}^{2}\\
+\frac{1}{8}\Tr\left[B_{k_1}^n h_{i_1}^{2n}\right]
 \Tr\left[B_{k_1}^n h_{j_1}^{2n}\right] c_{i_1,j_1,l_1}^{2n}
 \Tr\left[C_{k_2}^1 h_{i_2}^{2}\right]
 \Tr\left[C_{k_2}^1h_{j_2}^{2}\right] b_{i_2,j_2,l_2}^{2}\\
+\frac{1}{8}\Tr\left[C_{k_1}^n h_{i_1}^{2n}\right]
 \Tr\left[B_{k_1}^n h_{j_1}^{2n}\right] b_{i_1,j_1,l_1}^{2n}
 \Tr\left[B_{k_2}^1 h_{i_2}^{2}\right]
 \Tr\left[C_{k_2}^1h_{j_2}^{2}\right] c_{i_2,j_2,l_2}^{2}\\
+\frac{1}{8}\Tr\left[B_{k_1}^n h_{i_1}^{2n}\right]
 \Tr\left[C_{k_1}^n h_{j_1}^{2n}\right] b_{i_1,j_1,l_1}^{2n}
 \Tr\left[C_{k_2}^1 h_{i_2}^{2}\right]
 \Tr\left[B_{k_2}^1h_{j_2}^{2}\right] c_{i_2,j_2,l_2}^{2}.
 \label{eq:chchc2}
\end{multline}
For the recurrence relation \eqref{eq:chchc2} to be complete
three more terms are needed:
$\Tr\left[B_{k_1}^n h_{i_1}^{2n}\right]
 \Tr\left[B_{k_1}^n h_{j_1}^{2n}\right] c_{i_1,j_1,l_1}^{2n}$,
$\Tr\left[C_{k_1}^n h_{i_1}^{2n}\right]
 \Tr\left[B_{k_1}^n h_{j_1}^{2n}\right] b_{i_1,j_1,l_1}^{2n}$ and
$\Tr\left[B_{k_1}^n h_{i_1}^{2n}\right]
 \Tr\left[C_{k_1}^n h_{j_1}^{2n}\right] b_{i_1,j_1,l_1}^{2n}$.
Following the same
pathway as for Eq. \eqref{eq:chchc2}, we obtain
that the remaining recurrence relations are
\begin{multline}
\Tr\left[B_k^{n+1} h_i^{2(n+1)}\right]
\Tr\left[B_k^{n+1} h_j^{2(n+1)}\right]
c_{i,j,l}^{2(n+1)}
=\frac{1}{8}\Tr\left[C_{k_1}^n h_{i_1}^{2n}\right]
 \Tr\left[C_{k_1}^n h_{j_1}^{2n}\right] c_{i_1,j_1,l_1}^{2n}
 \Tr\left[C_{k_2}^1 h_{i_2}^{2}\right]
 \Tr\left[C_{k_2}^1h_{j_2}^{2}\right] b_{i_2,j_2,l_2}^{2}\\
+\frac{1}{8}\Tr\left[B_{k_1}^n h_{i_1}^{2n}\right]
 \Tr\left[B_{k_1}^n h_{j_1}^{2n}\right] c_{i_1,j_1,l_1}^{2n}
 \Tr\left[B_{k_2}^1 h_{i_2}^{2}\right]
 \Tr\left[B_{k_2}^1h_{j_2}^{2}\right] b_{i_2,j_2,l_2}^{2}\\
-\frac{1}{8}\Tr\left[C_{k_1}^n h_{i_1}^{2n}\right]
 \Tr\left[B_{k_1}^n h_{j_1}^{2n}\right] b_{i_1,j_1,l_1}^{2n}
 \Tr\left[C_{k_2}^1 h_{i_2}^{2}\right]
 \Tr\left[B_{k_2}^1h_{j_2}^{2}\right] c_{i_2,j_2,l_2}^{2}\\
-\frac{1}{8}\Tr\left[B_{k_1}^n h_{i_1}^{2n}\right]
 \Tr\left[C_{k_1}^n h_{j_1}^{2n}\right] b_{i_1,j_1,l_1}^{2n}
 \Tr\left[B_{k_2}^1 h_{i_2}^{2}\right]
 \Tr\left[C_{k_2}^1h_{j_2}^{2}\right] c_{i_2,j_2,l_2}^{2},
\end{multline}
\begin{multline}
\Tr\left[C_k^{n+1} h_i^{2(n+1)}\right]
\Tr\left[B_k^{n+1} h_j^{2(n+1)}\right]
b_{i,j,l}^{2(n+1)}
=-\frac{1}{8}\Tr\left[C_{k_1}^n h_{i_1}^{2n}\right]
 \Tr\left[C_{k_1}^n h_{j_1}^{2n}\right] c_{i_1,j_1,l_1}^{2n}
 \Tr\left[B_{k_2}^1 h_{i_2}^{2}\right]
 \Tr\left[C_{k_2}^1h_{j_2}^{2}\right] c_{i_2,j_2,l_2}^{2}\\
-
\frac{1}{8}\Tr\left[B_{k_1}^n h_{i_1}^{2n}\right]
 \Tr\left[B_{k_1}^n h_{j_1}^{2n}\right] c_{i_1,j_1,l_1}^{2n}
 \Tr\left[C_{k_2}^1 h_{i_2}^{2}\right]
 \Tr\left[B_{k_2}^1h_{j_2}^{2}\right] c_{i_2,j_2,l_2}^{2}\\
+\frac{1}{8}\Tr\left[C_{k_1}^n h_{i_1}^{2n}\right]
 \Tr\left[B_{k_1}^n h_{j_1}^{2n}\right] b_{i_1,j_1,l_1}^{2n}
 \Tr\left[B_{k_2}^1 h_{i_2}^{2}\right]
 \Tr\left[B_{k_2}^1h_{j_2}^{2}\right] b_{i_2,j_2,l_2}^{2}\\
+\frac{1}{8}\Tr\left[B_{k_1}^n h_{i_1}^{2n}\right]
 \Tr\left[C_{k_1}^n h_{j_1}^{2n}\right] b_{i_1,j_1,l_1}^{2n}
 \Tr\left[C_{k_2}^1 h_{i_2}^{2}\right]
 \Tr\left[C_{k_2}^1h_{j_2}^{2}\right] b_{i_2,j_2,l_2}^{2},
\end{multline}
and
\begin{multline}
\Tr\left[B_k^{n+1} h_i^{2(n+1)}\right]
\Tr\left[C_k^{n+1} h_j^{2(n+1)}\right]
b_{i,j,l}^{2(n+1)}
=\frac{1}{8}\Tr\left[C_{k_1}^n h_{i_1}^{2n}\right]
 \Tr\left[C_{k_1}^n h_{j_1}^{2n}\right] c_{i_1,j_1,l_1}^{2n}
 \Tr\left[C_{k_2}^1 h_{i_2}^{2}\right]
 \Tr\left[B_{k_2}^1h_{j_2}^{2}\right] c_{i_2,j_2,l_2}^{2}\\
-\frac{1}{8}\Tr\left[B_{k_1}^n h_{i_1}^{2n}\right]
 \Tr\left[B_{k_1}^n h_{j_1}^{2n}\right] c_{i_1,j_1,l_1}^{2n}
 \Tr\left[B_{k_2}^1 h_{i_2}^{2}\right]
 \Tr\left[C_{k_2}^1h_{j_2}^{2}\right] c_{i_2,j_2,l_2}^{2}\\
-\frac{1}{8}\Tr\left[C_{k_1}^n h_{i_1}^{2n}\right]
 \Tr\left[B_{k_1}^n h_{j_1}^{2n}\right] b_{i_1,j_1,l_1}^{2n}
 \Tr\left[C_{k_2}^1 h_{i_2}^{2}\right]
 \Tr\left[C_{k_2}^1h_{j_2}^{2}\right] b_{i_2,j_2,l_2}^{2}\\
+\frac{1}{8}\Tr\left[B_{k_1}^n h_{i_1}^{2n}\right]
 \Tr\left[C_{k_1}^n h_{j_1}^{2n}\right] b_{i_1,j_1,l_1}^{2n}
 \Tr\left[B_{k_2}^1 h_{i_2}^{2}\right]
 \Tr\left[B_{k_2}^1h_{j_2}^{2}\right] b_{i_2,j_2,l_2}^{2}.
 \label{eq:bhchb2}
\end{multline}
\end{widetext}
By recursively inputing \eqref{eq:recvanish}
into the
four recurrence relations
\eqref{eq:chchc2}-\eqref{eq:bhchb2}
it is readily verified that
\eqref{eq:chchc2}-\eqref{eq:bhchb2} vanish
for any number $n$ of qbits .
In particular, the fact that \eqref{eq:chchc2}
is zero proves that the projections
$\Tr[C_k^n h_i^{2n}]h_i^{2n}$ and
$\Tr[C_k^n h_j^{2n}]h_j^{2n}$
in Eq. \eqref{eq:commu0} indeed commute,
which is an essential factor
in expressing the Hamiltonian gates
of Eq. \eqref{eq:hamgate3}
as a succession of Pauli gates .

%

\end{document}